
%
%
%
%
%

\input jnl.tex
\input reforder.tex
\input eqnorder

\def\phys{Phys. Rev. B}
\def\lett{Phys. Rev. Lett.}


\def\dx2y2{\rm d_{x^2-y^2}}
\def\ds{\rm D_s}

\def\x214x{\rm La_{2-x} Sr_x Cu O_4}
\def\s2212{\rm Bi_2 Sr_2 Ca Cu_2 O_8}

\singlespace

\null
\vskip -.75in
\vskip .5in
{\singlespace
\smallskip
\rightline{ September, 1993}
}

\vskip 0.1in

\title SUPERCONDUCTIVITY NEAR PHASE SEPARATION
       IN MODELS OF CORRELATED ELECTRONS

\vskip 0.3in

\author  E.~DAGOTTO, J.~RIERA$^{1}$, Y.~C.~CHEN$^{2}$,
         A.~MOREO, A.~NAZARENKO, F.~ALCARAZ$^{3}$, and F.~ORTOLANI$^{4}$

\vskip 0.3cm

\affil Department of Physics,
       National High Magnetic Field Laboratory,
       and MARTECH,
       Florida State University,
       Tallahassee, FL 32306

\vskip 0.1in

\singlespace

\abstract {
\singlespace
Numerical and analytical studies of several models of correlated electrons
are discussed.
Based on exact diagonalization and variational Monte Carlo techniques,
we have found strong indications that the two dimensional
${\rm t-J}$ model superconducts near phase separation in the regime of
quarter-filling density, in agreement with previous results reported by
Dagotto and Riera (Phys. Rev. Letters ${\bf 70 }$, 682  (1993)).
At this density the dominant channel is $\dx2y2$, but
a novel transition to $s$-wave superconductivity is observed decreasing
the electronic density. In addition, the one band ${\rm t-U-V}$
model has also been studied using
the mean-field approximation that accurately described the spin density
wave phase of the repulsive Hubbard model at half-filling. An interesting
region of
$\dx2y2$ superconductivity near phase separation is observed in the
phase diagram,
and its implications
for recent self-consistent studies of d-wave condensates in the context
of the high-Tc cuprates are briefly discussed. Finally, the two band Hubbard
model on a chain
is also analyzed.
Superconducting correlations near phase
separation exist in this model, as
it occurs in the ${\rm t-J}$ model. Based on these nontrivial examples
it is
$conjectured$ that electronic models tend to have
superconducting phases in the vicinity of phase separation, and this regime
of parameter space should be explored first when a new model for
superconductivity is
proposed. Reciprocally, if it is established that a
model does not phase separate, even in an extended parameter space, then
we believe that
its chances of presenting a superconducting phase  are considerably reduced.
}

\vskip 0.4truecm

\line{PACS Indices: 75.10.Jm, 75.40.Mg, 74.20.-z\hfill }

\vskip 1truecm

\singlespace

{\singlespace
\noindent
1) { Permanent address:
Center for Computationally Intensive Physics,
Physics Division, Oak Ridge National Laboratory,
Oak Ridge, Tennessee 37831-6373
and
Department of Physics $\&$ Astronomy,
Vanderbilt Univ.
Nashville, Tennessee 37235.} }

{\singlespace
\noindent
2) { Permanent address: Department of Physics,
National Tsing Hua University, Hsinchu 30043, Taiwan, R.O.C. }
}

{\singlespace
\noindent
3) { Permanent address: Departamento de Fisica,
Universidade Federal de Sao Carlos, CP 616, 13560,
Sao Carlos, SP, Brasil. }
}

{\singlespace
\noindent
4) { Permanent address: Dipartimento di Fisica, Instituto Nazionale
di Fisica Nucleare, Universit\`a di Bologna, via Irnerio 46, I-40126
Bologna,
Italy.}
}


\endtitlepage

\singlespace

\vskip 1.cm

\centerline{\bf I. Introduction}

\vskip 0.5cm

The study of high temperature superconductors continues attracting
much attention.\refto{muller} On the
experimental side, steady progress is being made in the preparation of
single crystal samples of high quality for several compounds.
The common features of the cuprates are experimentally well
established, specially the ``anomalous'' behavior of several
observables in the normal state like the d.c. resistivity, optical
conductivity $\sigma(\omega)$, Hall coefficient, and
others.\refto{batlogg}
New compounds have been recently discovered with a critical temperature
of $\sim 133K$, increasing our expectations that the cuprates may become
technologically relevant in the near future.\refto{schilling}
On the theoretical
side, also considerable progress has been achieved in recent years
in the study of models of correlated electrons. Powerful numerical
techniques have shown that some of the anomalous properties of
the cuprates may be explained by simple one band Hubbard and ${\rm t-J}$
models.\refto{review,dago92}
The mid-infrared band of $\sigma(\omega)$ observed in $\x214x$ and
other compounds\refto{uchida} may be
produced, at least in part, by the spin excitations that heavily dress the
hole carriers.\refto{review}
The temperature dependence of the magnetic susceptibility\refto{dcj}
can also be explained by models of holes moving in a strong
antiferromagnetic
background.\refto{moreo93} The appearance of states in the charge transfer gap
upon doping\refto{fujimori}
can be mimicked using the one band Hubbard model.\refto{prl91}
Mean-field theories of the ${\rm t-J}$ model\refto{fukuyama}
have contributed
to the understanding
of the incommensurate peaks
observed in $\x214x$,\refto{cheong}
and they are in good agreement with numerical
studies of the same model.\refto{moreo}
Unfortunately, other anomalous properties are still
unexplained, like the linear behavior of the resistivity with temperature, and
the pinning of the chemical potential with doping in some
compounds.\refto{allen}
Nevertheless, the theoretical progress described above should be considered as
an
important step towards the development of a microscopic theory of
the high-Tc cuprates.

In spite of this progress in the description of normal state properties,
the presence of superconductivity in the
ground state of models of correlated electrons is still a subtle and
much debated issue.
Numerical studies of the ${\rm t-J}$ model on large enough lattices have
shown that holes in an antiferromagnetic background tend to form bound
states in the $\dx2y2$ channel.\refto{binding}
Similar results have been obtained
in the context of the one band Hubbard model using
self-consistent approximations which suggest the existence of d-wave
pairing at very low temperatures.\refto{bulut}
However,
no convincing Quantum Monte Carlo numerical
evidence has been found that the model indeed superconducts at low
temperatures and hole density.\refto{moreo92} These numerical results
can be interpreted in two ways. On one hand, it may occur
that the Hubbard model indeed does not
superconduct. This negative result is not al all excluded.
On the other hand, it may occur that present day numerical
techniques are not accurate enough to find out the (weak) signals of a
superconducting condensate in this model. It is clear that analyzing a two
hole problem is simpler than searching for a condensate, where the
subtle coherence and overlap  effects between pairs is crucial for its
existence.
In particular note that for a finite lattice of, e.g., 16 sites, a
realistic electronic density of $\langle n \rangle = 0.875$ corresponds
to only one pair of holes, which of course cannot produce a superconducting
signal alone. Then, it may simply occur that close to half-filling the
clusters accessible to numerical studies do not have enough pairs to
produce a clear signal of superconductivity as an output.
Actually, Monte Carlo studies of the attractive Hubbard
model\refto{scalettar} (which has a well-established
superconducting ground state) have shown that in the regime of low electronic
density (and thus low pair density) it is difficult to observe a numerical
evidence of
superconductivity
in the ground state. A similar situation may occur in the ${\rm t-J}$ and
repulsive Hubbard models near half-filling where the density of pairs is small.

Based on these ideas, two of us\refto{prl} recently started the search for
indications of superconductivity in the ground state of the two
dimensional ${\rm t-J}$ model in a novel
regime of parameter space, namely $\langle n \rangle = 1/2$ (which by analogy
with the Hubbard model, it will be called ``quarter-filling'' in this
paper),
and large coupling ${\rm J/t}$. The main motivation for such a study
in a region of parameter space which is not realistic for the cuprates
is that in this regime the number of carrier pairs is maximized (closer to
half-filling there are fewer holes, and near the empty system fewer
electrons). In addition, it is well-known that there are attractive forces
operating in this model since the system phase separates at large ${\rm
J/t}$. Also note that
at low electronic density, two
electrons on an otherwise empty lattice
minimize their energy by forming a tight singlet bound state.\refto{phsep}
Then, densities of quarter-filling or less and couplings close to
phase separation seem the most optimal regime to search for
superconductivity in the two dimensional (2D) ${\rm t-J}$ model, as
reported by Dagotto and Riera.\refto{prl}
A strong indication of superconductivity was
found by these authors in
the equal-time pairing correlations, with a signal which is maximized
in the $\dx2y2$ channel in agreement with the results for two holes
close to half-filling,\refto{binding} and with the self-consistent
approximations.\refto{bulut}

One of the purposes of this paper is to discuss in more detail the
results found in Ref.(\cite{prl}) for the 2D ${\rm t-J}$ model. We provide
numerical evidence showing that the region of d-wave superconductivity
near quarter-filling is robust and likely to survive the bulk
limit.
In addition, we found a novel transition from $\dx2y2$-wave to $s$-wave
superconductivity as a function of density. This transition was
expected in order to make compatible the results of
Emery, Kivelson and Lin\refto{phsep} suggesting s-wave superconductivity
at very low density (based on the presence of two electron s-wave bound
states on an empty lattice), and those of Dagotto and Riera\refto{prl} at
quarter filling, suggesting a $\dx2y2$ condensate.
The numerical
evidence for these results is based on a Variational Monte Carlo
calculation on large clusters, and it is consistent with the results
obtained with the Lanczos approach on smaller clusters. The success of
this search for superconductivity in the ${\rm t-J}$ model
opens the
possibility for the existence of superconductivity in the realistic
regime of small ${\rm J/t}$ and $\langle n \rangle \sim 1$, as discussed below
in the text.\refto{qcd}

Actually,
this paper has a more general purpose. We will argue that for a given
electronic model, the region in parameter space
where superconductivity has the highest
chances of being stable is
in the neighborhood of phase separation.\refto{grilli} Moreover, we
believe that
the attractive force responsible for phase separation also leads to
the strong pairing correlations in its neighborhood (similar ideas have
been recently discussed by Emery and Kivelson, see Ref.(\cite{emery93}) ). In
this
regime
it is energetically favorable to form pairs of holes rather than larger
clusters of holes, due to the gain in kinetic energy obtained by giving
mobility to those pairs.
Reciprocally, if the model
does not have phase separation, not even in an extended parameter space
(without explicitly
changing the repulsive or attractive character of the potential),
then it is difficult to imagine that it will present a superconducting
phase. This is a $conjecture$ that is presented and discussed in this paper.
As with any conjecture, we do not have a rigorous mathematical proof of its
validity, but here we provide several examples that illustrate the main
ideas behind it. This conjecture has practical implications since it is
well-known that establishing the presence of phase separation for a
given
model is usually simpler than finding a superconducting ground state.
Then,
once phase separation is found, the region to  search for pairing is
considerably reduced. To support our ideas,
in addition to the two dimensional ${\rm t-J}$ model,
in this paper we also discuss the attractive Hubbard model which naively seems
a
counter-example to our conjecture i.e. it superconducts but does not
phase separate. However, it can be shown that in an extended
parameter space (i.e. including attractive density-density correlations which
are usually spontaneously generated by any renormalization group
procedure), the regime of s-wave superconductivity of the negative U
Hubbard model is in contact with a robust region of phase separation.
As a bonus of this study,  an interesting region of $\dx2y2$
superconductivity in the extended ${\rm t-U-V}$ one band Hubbard model was
observed. This phase
may have important implications for recent efforts to describe
the cuprates with effective Hamiltonians based on the interchange of
magnons.\refto{bulut} Finally, we briefly present results for the two
bands Hubbard model in a one dimensional
chain showing that also in this case strong
superconducting correlations appear near phase
separation. All these examples give strong support to our conjecture.
In addition, note that recent experimental work\refto{phsepexp} has shown that
in some cuprates with an excess of oxygen there is phase separation
and is not caused just by a chemistry problem (the oxygens are very
mobile). Theoretical ideas by Emery and Kivelson\refto{emery93}
have been presented linking phase
separation and superconductivity in the cuprates, and
thus the type of studies described in this paper may have implications
for real materials.

The organization of the paper is the following: in section II the
t-J model in two dimensions is studied at different densities. The
interesting transition from $\dx2y2$-wave to $s$-wave superconductivity
is discussed. In section III, mean field results
for the $t-U-V$ model are presented, and its rich phase diagram
discussed at half-filling. In section IV, we briefly discuss results for the
two band Hubbard model on a chain. Conclusions are presented in section
V, and finally in the appendices we discuss measurements of the
superfluid
density for the one band Hubbard model, and the influence of the
fermionic statistics on the phase diagram of the 2D ${\rm t-J}$ model.

\vskip 1.9cm

\centerline{\bf II. ${\rm t-J}$ model in two dimensions}

\vskip 0.5cm

\noindent{\bf a) Density $\langle n \rangle = 1/2$}

\vskip 0.2cm

As discussed in the Introduction, recently it has been suggested by two of
us\refto{prl} that the ${\rm t-J}$ model in two dimensions may have
a superconducting phase near phase separation, and at density $\langle n
\rangle = 1/2$. The numerical evidence for these conclusions was based
on an exact diagonalization study of finite clusters, analyzing the
equal-time pairing correlations, the superfluid density (discussed
briefly in Appendix A), and the
anomalous flux quantization. For completeness, in this section some
of those results are reproduced from Ref.(\cite{prl}).
New information for other cluster sizes and
parameters is provided.
The ${\rm t-J}$ model is defined by the Hamiltonian,
$$
{\rm H =
{\rm J } \sum_{{\bf \langle i j\rangle }}
( {{\bf S}_{\bf i}}.{{\bf S}_{\bf j}} - {1\over4} n_{\bf i} n_{\bf j} )
- {\rm t} \sum_{{\bf \langle i j \rangle},s}
({\bar c}^{\dagger}_{{\bf i},s} {\bar c}_{{\bf j},s} + h.c.) },
\tag 1
$$
\noindent where
${\rm  {\bar c}^{\dagger}_{{\bf i},s}}$
denote $hole$ creation operators;
${\rm n_{\bf i} = n_{{\bf i},\uparrow} + n_{{\bf i},\downarrow} }$ are
number
operators for electrons;
and the rest of the notation is standard.
In the numerical studies of this model described below, square clusters
of ${\rm N}$ sites
have been used (with periodic boundary conditions).
In order to search for indications of superconductivity in the ground state,
let us define the pairing operator
$\Delta_{\bf i} = {c}_{{\bf i},\uparrow}
(  {c}_{{\bf i+{\hat x}},\downarrow} +
   {c}_{{\bf i-{\hat x}},\downarrow} \pm
   {c}_{{\bf i+{\hat y}},\downarrow} \pm
   {c}_{{\bf i-{\hat y}},\downarrow} )$,
where $+$ and $-$ corresponds to
extended-s and $\dx2y2$ waves, respectively,
and ${\bf {\rm {\hat x},{\hat y}}}$ are unit vectors along the axis.
Note that the pairing operator is constructed using electronic operators
(not hole operators).
The pairing-pairing correlation function defined as
${\rm C({\bf m}) = {1\over N} \sum_{{\bf i}} \langle \Delta^{\dagger}_{\bf i}
\Delta_{{\bf i + m}} \rangle}$,
and its susceptibility $\chi^\alpha_{sup}
= {\rm \sum_{{\bf m}} C({\bf m})}$ have been calculated
(where $\alpha = d$ corresponds to
$\dx2y2$ wave, and $\alpha = s$ to extended-s wave).
$\langle \rangle$ denotes the expectation value in the ground state, which
is obtained accurately using the Lanczos method.
Note that the pairing operator used here is not strictly a spin
singlet but actually the sum of a singlet and a triplet.
This operator
allows us to study the two spin sectors at the same time. Once a large signal
in
the pairing correlations is
observed in some region of parameter space, it is trivial to implement
rigorous spin singlet or triplet operators to find out which one carries
the strongest correlation. In practice, we found that the results
obtained using the operator $\Delta_{\bf i}$ defined above
at all distances
larger than zero, are
quantitatively
very similar to the results obtained
using a spin-singlet operator. Even at zero distance the difference is only of
about $15\%$.
Then, the results below using $\Delta_{\bf i}$ should be considered
approximately
equal to those
obtained using the proper singlet operator $\Delta^S_{\bf i}$ ( where
$\Delta^S_{\bf i}  = \Delta_{\bf i} -  {\hat I} \Delta_{\bf i}$, and
${\hat I}$ is the operator that inverts the z-direction spin projection
for each electron).

Working on a $4 \times 4$ cluster, and density $\langle n \rangle =
1/2$, the numerical results are shown in Fig.1.
The d-wave susceptibility presents a sharp peak at coupling ${\rm
J/t = 3}$, suggesting that strong pairing correlations exist in
this region of parameter space (Fig.1a). We have verified
explicitly that the abrupt
reduction of the signal after ${\rm J/t=3}$ is caused by a transition
to the phase separation region which is expected at large
couplings\refto{phsep}
(this conclusion was obtained studying numerically the compressibility).
However, it is important to remark that a large superconducting
susceptibility (or rather, zero
momentum pairing correlation function) is not sufficient to guarantee
the presence of long-range order, but an increase of $\chi^d_{sup}$ with
the lattice size is required. Unfortunately, it is
difficult to study lattices much larger than the $4 \times 4$ cluster,
and thus such an explicit analysis is not possible with present day
computers.
Nevertheless, we can study the existence of long-range order in this model by
explicitly calculating the pairing correlations as a function of distance.
A robust tail in the correlation would suggest long-range order (or at
least
a correlation length larger than the lattice size). The
results for the ${\rm t-J}$ model are shown in Fig.1b, both for the
$\dx2y2$ and extended-s wave correlations. Similar
correlations are also shown in Fig.1c for
the d-wave channel, parametric with the coupling ${\rm J/t}$. These
results indicate that most of the signal in the susceptibility does $not$
come from the on-site correlation, but from its tail. Although this is
not a rigorous proof, such a result strongly suggests that long distance
pairing correlations are developing in this region of parameter space.
Further studies at other densities and couplings show that the
results of Fig.1b-c are indeed robust, and in regions where no
superconductivity exists, the pairing correlations decay very
abruptly with distance, typically being compatible with zero at about
two lattice spacings. This is precisely the case of the repulsive Hubbard model
in two dimensions studied with Quantum Monte Carlo
techniques.\refto{moreo92} Then,
the reader should notice that the correlations shown in
Fig.1b-c are perhaps the
strongest numerical signals of superconductivity reported thus far in the
literature of the 2D ${\rm t-J}$ and Hubbard models, using unbiased
numerical techniques.


What is the physical reason for the presence of strong pairing
correlations in this region of parameter space?
To begin with, note that there are attractive forces between
electrons operative in the ${\rm t-J}$ model which are responsible for the
existence of phase separation. Such forces can be roughly
described in the two limits of low and high electronic density. For
example, it is well known that two holes in an antiferromagnetic
background at large ${\rm J/t}$ tend to bind in a $\dx2y2$-wave bound
state in order to minimize the number of antiferromagnetic missing
bonds.\refto{binding}
This force leads to clustering of holes at a finite hole density. In
the other limit of low electronic density, it can be shown that two electrons
in an otherwise
empty lattice form a bound state at coupling ${\rm J/t > 2}$, since
the Heisenberg term acts like an explicitly attractive force. Increasing
further the coupling and working at finite density,
phase separation occurs.\refto{phsep}
This special case is interesting since it shows that at low electronic
density there are bound states whose kinetic energy forbids the
clustering effect until a large coupling is reached. We believe that
a similar scenario holds at quarter filling $\langle n \rangle = 1/2$,
namely that the force that produces phase separation for ${\rm J/t >3}$
also produces pairing at smaller coupling. The gain in
kinetic energy of the mobile
pairs forbids phase separation in this regime.

Another qualitative argument to help understanding the presence of
pairing is the following: suppose a repulsive density-density interaction
${\rm V \sum_{\langle ij \rangle} n_i n_j}$ is added to the Hamiltonian
of the ${\rm t-J}$ model. Such a term has been analyzed by Kivelson,
Emery, and Lin\refto{kivelson} in the large ${\rm V/t}$ limit, and by
Dagotto and
Riera\refto{super}
numerically for all values of ${\rm V/t}$. These authors showed that
at large ${\rm V/t}$ a tendency
to form an ordered arrangement of dimers
(spin singlets formed by two electrons at a distance of one lattice
spacing) exists.
Analytic calculations, supported by numerical results, show
the existence of this dimer lattice very clearly. The introduction of
a hopping term $t$ induces superconducting correlations.\refto{kivelson}
The exact diagonalization results\refto{super} suggest a smooth connection
between
large and small ${\rm V/t}$, and thus some remnants of dimers may
exist in the pure ${\rm t-J}$ model at this density (although now in a
spatially disordered state).
The phase diagram obtained numerically
in the plane ${\rm V/t-J/t}$ is shown in
Fig.2 (the points represent the position of the maximum in the superconducting
susceptibility evaluated on a $4 \times 4$ cluster).
However, a more subtle issue forbids a completely
smooth connection between the two regimes: at large ${\rm V/t}$ there is
a robust spin-gap in the spectrum due to the formation of dimers, and
thus the s-wave correlations dominate.
However, at ${\rm V/t=0}$ the $\dx2y2$ correlations are dominant and
the presence of nodes in the spectrum (in the bulk limit) allow
for the possibility of creating low
energy spin triplets (i.e. zero spin-gap). Then, if our numerical
results are valid in the bulk limit, a transition between s and d
wave ground states should exist as a function of ${\rm V/t}$, near
phase separation (the verification of this idea certainly deserves more work).
The closing of the spin gap as a function of this
coupling has indeed been observed by Troyer et al. in the one dimensional
version of the ${\rm t-J-V}$ model.\refto{troyer} In the
next section, it will be shown that a transition from d to s-wave
condensates also exists as a
function of density in the pure 2D ${\rm t-J}$ model.

It would be important to study larger clusters in order to verify that
the strong tail of the correlations as a function of distance
shown in Fig.1b-c are not a mere finite size effect. Unfortunately,
such a study is difficult with present day computers since the memory
requirements needed to carry out a Lanczos study of the ${\rm t-J}$ at
quarter filling grow exponentially with the cluster size.\refto{trunca}
In spite of
this problem we managed to study a tilted cluster of 20 sites\refto{oitmaa} as
that
shown in Fig.3a. Although not obvious to the eye, it can be shown that
this cluster is
invariant under rotations in $\pi/2$ about a site, and thus it can
be used to explore the presence of superconductivity in the
$d$-channel.\refto{foot}
However, the 20 site cluster has a disadvantage for the particular
problem studied in this section. The trouble is schematically shown in
Fig.3a: considering the ${\rm t-J-V}$ model in the limit of
large ${\rm V/t}$, the five dimers that could in principle be formed in the
cluster
do not have space to minimize the energy unless a ${\rm V}$-energy larger than
${\rm 5V}$ is paid.
This is purely an artifact of the shape of the cluster, that does not
occur in square $L \times L$ lattices. Since on the $4 \times 4$
cluster a smooth connection was observed between large and small ${\rm V/t}$,
the lack of a proper ${\rm V/t}$ limit for the 20 site cluster implies
that the signal for superconductivity in the other limit of
${\rm V/t=0}$ may be
spuriously reduced compared to that of the 16 site cluster. This
conclusion was verified by an explicit numerical study of the N=20
cluster. The normalized signal for $d$-wave pairing
correlations shown in Fig.3b
is reduced by approximately a factor two when the lattice size is
increased from
N=16 to 20 sites. Rather than considering this as a negative result for our
scenario,
we believe the reason for this reduction is the topology of the N=20
cluster as explained before. In spite of this problem, note that even for the
N=20
lattice, there are no indications that the
correlation will decay to zero at large distances since the signal is
fairly flat. Of course, studies on larger clusters
that satisfy the proper ${\rm V/t}$ limit would be important to verify
our assumptions, but we believe that the evidence discussed in this
section suggesting the existence of superconductivity in the quarter-filled 2D
${\rm t-J}$ model is strong, and may survive the bulk limit.
Finally, we would like to remark that ``dynamical'' studies of this
condensate would be very important. Actually, we have already carried
out studies of the dynamical pairing correlation in the d-wave channel.
It shows a sharp peak at the bottom of the spectrum, as expected from a
d-wave condensate. Other dynamical properties are currently being
analyzed by S. Maekawa et al.\refto{ohta}

\vskip 0.5cm

\noindent{\bf b) Low electronic density, $\langle n \rangle < 1/2$}

\vskip 0.2cm

It is interesting to extend the results obtained in the previous
subsection to other densities. Here, the region $\langle n
\rangle < 1/2$ will be explored. In this regime the problem associated with the
lattices
of 20 sites does not hold anymore since four or less dimers can be
accommodated in the cluster of Fig.3a without trouble at large ${\rm
V/t}$,
and thus results
for a lattice slightly larger than a $4 \times 4$ cluster become available.
In addition, there is a physical motivation for the study of
low electronic densities: from the work of Emery, Kivelson and
Lin\refto{phsep,lin2}
it is known that
at ${\rm J/t=2}$ a bound state of two electrons appears due to the
attractive spin-spin interaction in the Hamiltonian that favors the
formation of a spin singlet state. Then, it is natural to expect that
a finite (but small) density of
these bound states may Bose condense at low temperatures in the s-wave
channel.\refto{phsep,kivelson} This argument is very persuasive, but the
numerical results in favor of a d-wave condensate
at quarter filling are also fairly strong.\refto{prl} Thus, the only solution
to
this apparent paradox is that the ground state of the 2D ${\rm t-J}$ model
exhibits a transition from s-wave at low densities to
d-wave at higher densities. Here, evidence based on
exact diagonalization and variational Monte Carlo (VMC) studies is
presented to
support this conjecture.\refto{tklee} Then, the phase diagram of the ${\rm
t-J}$ model
seems very rich indeed showing d and s-wave superconducting condensates,
phase separation, antiferromagnetism, ferromagnetism, and paramagnetic
phases in the $ \langle n \rangle$-${\rm J/t}$ plane.

Consider a 20 site cluster with 8 electrons (i.e. $\langle n
\rangle = 0.4$), and let us evaluate in its ground state the
same pairing correlation functions studied at quarter filling in
Fig.1b-c. The results for the $\dx2y2$ and extended s symmetries are
shown in Fig.4a,b. At this density, it is clear from the figure that
the d-wave pairing correlations are still dominant over s-wave, i.e. the
d-wave correlations at a distance of approximately three lattice spacings
are robust and do not show indications of reduction with distance.
In Fig.4c,d similar correlations are shown at a lower density
$\langle n \rangle = 0.2$ ( 4 electrons on the 20 site cluster). The
qualitative results are similar to those obtained in Fig.4a
although now the correlation at a distance of two lattice spacings
in the s-wave channel
is larger than at $\langle n \rangle = 0.4$. This suggests a tendency
towards the formation of a competing s-wave condensate, but it is not enough
to show that such a condensate will become stable upon further reduction
of the density. Unfortunately, on this cluster the next density
available corresponds to only two electrons which we know cannot be
representative of a finite density of particles. Thus, from the exact
diagonalization analysis we can only roughly say that the transition from d to
s-wave superconductivity may occur at an electronic density $\langle n \rangle
< 0.2$.

It would be very important to verify this result by some other
independent calculation. For this purpose,
we have implemented a simple VMC simulation, using trial wave functions with
an s and d symmetry superconducting condensates,
and also states representing a Fermi liquid, and the phase
separated regime. From the energy competition between these states, we should
be able to
extract qualitative information at low densities. The states we
have used are a Gutzwiller state, which becomes stable at small
values of ${\rm J/t}$, and is defined as,
$$
| GW \rangle = \sum_{\bf r_1,...,r_{N_e}} P_d\quad det_\uparrow \quad
det_\downarrow
 \quad \Pi_{ {\bf r_{i}} \sigma_i } c^\dagger_{ {\bf r_i}\sigma_{\bf i}} | 0
\rangle,
\tag {s4}
$$
\noindent where
$P_d = \Pi_{\bf i}    (1 - n_{{\bf i}\uparrow} n_{{\bf i}\downarrow} )$
projects out states with double occupancy, $det_\sigma$ are Slater
determinants of a filled Fermi sea  corresponding to spin $\sigma$, and
the sum in front denotes all possible electronic configurations (with the spin
index omitted). ${\rm N_e}$ is the number of electrons,
while the rest of the notation is standard, and
follows the recent work of Valenti and Gros on variational wave
functions.\refto{valenti}

As superconducting condensates we use the states
previously introduced by Gros et al.\refto{gros} For the $\alpha$-wave
condensate
($\alpha$ = s or d) we define,
$$
| \alpha \rangle \sim P_d ( \sum_{\bf k} a_{\bf k}
c^\dagger_{{\bf k}\uparrow} c^\dagger_{{\bf -k}\downarrow} )^{\rm N_e/2} | 0
\rangle,
\tag {s5}
$$
\noindent where $a_{\bf k} = \Delta_{\bf k} / ( \epsilon_{\bf k} +
\sqrt{ \epsilon^2_{\bf k} + \Delta^2_{\bf k} } )$, and
$\epsilon_{\bf k} = -2t (cosk_x + cosk_y) - \mu$ ($\mu$ is the chemical
potential).
The parameter $\Delta_{\bf k} =
\Delta_V$
corresponds to s-wave, while $\Delta_{\bf k} =
\Delta_V (cosk_x - cosk_y) $ is a $\dx2y2$ wave. $\Delta_V$ is a ${\bf
k}$-independent variational
parameter (note that this state can be
rearranged in the form of a RVB state, see Ref.(\cite{gros})).
Finally, a variational state for the phase separated region was used, which
simply has all the electrons clustered in an antiferromagnetic state
(whose energy density can be easily obtained from Monte Carlo
calculations of the spin-1/2 Heisenberg model in two dimensions\refto{barnes}).

The actual variational calculations have been carried out using the
Monte Carlo technique on an $8 \times 8$ cluster, and for 10, 26, 42,
and 50 electrons (that correspond to closed shell configurations).
About 10,000 Monte Carlo sweeps for each density and coupling were
performed. The results for the energy of each one of these variational
states are shown in Fig.5a-c. At low electronic density $\langle n
\rangle = 0.156$, Fig.5a shows
that the s-wave state is energetically better than the corresponding
d-wave state approximately in
the interval ${\rm 4 < J/t < 5.5 }$. For smaller values of the coupling,
the Gutzwiller state dominates, while for larger values of ${\rm J/t}$
the phase separated state is stable, as expected.
The dominance of s-wave correlations near phase separation at low fillings,
is compatible with the ideas discussed before, namely that electrons
in an otherwise empty lattice bound in s-wave states, and thus at least
for $\langle n \rangle << 1$ and low temperatures
we would expect a Bose condensation of these s-wave pairs.\refto{phsep,rpa}

Fig.5b shows results for a higher electronic density
$\langle n \rangle \sim 0.406$.
In this case, the s-wave is no longer stable, and the d-wave minimizes the
energy in the region ${\rm 2.5 < J/t < 3.8}$. This result is
to be expected based on our exact diagonalization analysis near quarter
filling. Then, as a function of density a transition from s-wave to
d-wave pairing has been observed. Increasing further the filling to $\langle
n \rangle = 0.656$, the d-wave state still dominates between
${\rm 1.5 < J/t < 3.0}$.
A rough and qualitative phase diagram obtained with the VMC method
is shown in Fig.5d. Results at densities close to half-filling are not
shown since the variational energies for different states
are very close to each other in this region, and thus
corrections to each state cannot be neglected.

\vskip 1cm

\noindent{\bf c) High electronic density, $\langle n \rangle > 1/2$:}

\vskip 0.3cm

After analyzing the region of small electronic density, it is important
to study the more realistic regime of high densities.
It is clear that the ${\rm t-J}$ model was originally introduced
to qualitatively mimic the behavior of the high-Tc cuprates, and thus
the physically interesting regime corresponds to small hole density
(i.e. $\langle n \rangle $ slightly smaller than one), and small values of
${\rm J/t}$ (since the cuprate compounds are known to be in the strong
coupling regime, and if the one band Hubbard Hamiltonian is used to
model the materials, we should work in the region ${\rm U/t >>1,}$ which
corresponds to ${\rm J/t << 1 }$).
Then, it is quite important to carry out a numerical study for densities
$\langle n \rangle > 1/2$. The results for 4 holes on
the 16 site cluster are shown in Fig.6a-b, where they are compared with
the results for 8 holes. The susceptibility at $\langle n \rangle =
0.75$ is much flatter than at quarter filling, and the actual correlations
as a function of distance are very close to zero already at a distance
of two lattice spacings. Then, our numerical results do not show
indications that the interesting superconducting
region observed at $\langle n \rangle \leq 1/2$ can be extended towards
the realistic regime of densities closer to half-filling.

However, we would like to point out that this negative result is not
definitive. There are many arguments suggesting that a calculation
similar to that of Fig.6a-b, but carried out on a larger lattice may
show more positive signals of superconductivity. To begin with, note
that the VMC calculation described in the previous section on a larger
lattice still predicts the dominance of a d-wave condensate over other
states. Although variational calculations are generally uncontrollable
(since it is difficult to judge how accurate the trial
states are), the interesting qualitative agreement found with the exact
diagonalization results at other densities suggests that the wave functions
we used may be quantitatively accurate.
The ``size'' in real space
of the pairing operators is another possible reason for the small
correlations found in Fig.6a-b. In a typical BCS condensate, the size of
a pair depends on the density of carriers, even with a strong local
attractive potential.
The lower the density, the larger the pair size to maintain the coherence
among pairs. Then, it may occur that the local operator used in our
study does not have a large overlap with the actual more extended
pairing operator
that may exist in this model.

Finally, we would like to point out an analogy between our results for
the 2D lattice, and those obtained for the 1D ${\rm t-J}$ chain. As it is
well-known, a numerical analysis of the ${\rm t-J}$ model in one dimension
has shown a region where superconducting correlations are dominant\refto{ogata}
in the sense that they decay the slowest as a function of distance.
This
region is a strip bordering phase separation (very similar to that found in our
two dimensional study), and it covers a wide range of densities
between small $\langle n \rangle$, and
$\langle n \rangle$ as large as $0.875$.
Although no long-range order can occur in 1D at any temperature,
and even the statistics of the particles
is not important in one dimension, since
fermions and hard core bosons produce the same phase diagram
(in two dimensions the situation is drastically different, see Appendix
B),
the results on a chain are still very instructive to guide our intuition in
the more realistic 2D problem.
The 1D results are based on a study of
$K_\rho$ which is a parameter based on conformal field theory,
that controls the decay of the correlation
functions with distance. This quantity can be obtained from a study
of spin and charge velocities, and it is believed to be affected by
finite size effects less severely than actual pair correlations.
In Fig.6c-d we show the pair susceptibility and correlation function
for the case of 8, 4 and 2 holes on a 16 site one dimensional chain. The
case of 2 holes corresponds to a nominal density $\langle n
\rangle = 0.875$ i.e. where superconductivity should still be dominant.
However, Fig.6d clearly shows that for this density an analysis of the
pair correlation shows no indications of the
superconductivity dominated ground state implied by $K_\rho$ (this is
not surprising since 2 holes can form only one pair). This example tells
us that a superconducting ground state cannot be identified easily
using finite clusters when only a few number of pairs is
available. It may occur that the regime of $\langle n
\rangle = 0.75$ in 2D is analog to $\langle n \rangle = 0.875$ in 1D.

Thus, we conclude that
due to the limitations of numerical studies based on exact
diagonalizations, the regime of small hole density is difficult to
analyze (unfortunately, there are no stable Quantum Monte Carlo techniques
to study the 2D ${\rm t-J}$ model at low temperatures). Then, we believe that
in order to make further progress closer to half-filling, it would be
important to develop a good variational wave function to describe the
regime of quarter filling (where results can be compared with the
exact diagonalization predictions), and then carry out calculations with
the same wave function at densities closer to half-filling.\refto{foot2}
Note also that other terms in the Hamiltonian like a ${\rm t'}$ hopping
may shift the d-wave region towards even smaller densities.

\vskip 1cm

\noindent{\bf d) Phase diagram of the 2D ${\rm t-J}$ model}

\vskip 0.3cm

Based on the numerical results discussed in the previous sections, we
believe that the phase diagram of the two dimensional ${\rm t-J}$
model is schematically as shown in Fig.7. Clearly there is a large region of
phase separation at large ${\rm J/t}$ for all values of the density.
The boundaries of this phase are in good agreement
with results from high temperature expansions.\refto{putikka}
The novel
result discussed in this paper, and before in Ref.(\cite{prl}), is
the existence of a region of d-wave superconductivity (in the
$\dx2y2$ channel) that extends from low to high electronic densities,
with a numerical signal that is maximized at quarter-filling (where
the numbers of pairs is also maximized). Numerically it is difficult
to find the boundary of the d-wave phase at large electronic densities,
i.e. close to half-filling. Even the phase separated region boundary
is controversial (and thus we only write ``AF'' in Fig.7 in that
regime, to show that antiferromagnetic correlations are important but
we do not know the details of the phase diagram).
However, we have presented arguments suggesting that
it may be possible that the d-wave region survives as a narrow strip
following phase separation, even in the regime close to half-filling.
After all, two holes in an antiferromagnetic background form a $\dx2y2$
bound state,\refto{binding}
and thus the most economical hypothesis is to link the
numerical strong signals at quarter-filling with the bound states at
half-filling. Of course, the verification of this hypothesis
needs, and deserves, more work.

In the other limit of low electronic density, an interesting change
in the symmetry of the condensate is observed (which was unknown to
two of us in a previous publication\refto{prl}).
This is compatible
with the observation that two electrons on an empty lattice form a
s-wave bound state for ${\rm J/t > 2}$.\refto{phsep}
This region of condensed dimers
may be similar to that found at large density-density repulsion
${\rm V/t >>1}$ in previous work.\refto{kivelson,super} We believe that in the
three dimensional parameter space ${\rm J/t, V/t, \langle n \rangle}$,
the s-wave region of the pure ${\rm t-J}$ model at low electronic
density, and the s-wave
region at quarter filling of the ${\rm t-J-V}$ model may be analytically
connected. A similar behavior may occur in one dimension where a
spin-gap
phase\refto{ogata,chen,hellberg}
was observed at low densities for the ${\rm t-J}$ model,
and at large ${\rm V/t}$ and
quarter-filling for the ${\rm t-J-V}$ model.\refto{kivelson,super} Such
a rich phase diagram certainly deserves more
work. The d-wave region found in this work
seems disconnected from those two limiting
cases with s-wave symmetry, and should be considered as a new phase.
Finally, the paramagnetic PM region resembles much a noninteracting gas of
particles (and thus a Fermi-liquid),
but a careful study of the wave function renormalization $Z$
is needed to clarify this issue.

It is important to remark that the region of s-wave superconductivity at
small electronic density shown
in Fig.7 may be larger that what is shown in the figure. In constructing
the boundary of phase separation in Fig.7 we used
the discrete version of the second derivative of the
ground state with respect to density. However, the error bars at small
$\langle n \rangle$ are large using this procedure. Then, in the region
$\langle n \rangle <<1$ we used the results obtained from the high
temperature expansions\refto{putikka} to complete the boundary of phase
separation. Although with this technique the study of pairing
correlations
has not been addressed yet, their predictions for the phase separation
boundary seem accurate at intermediate densities where the results can
be
contrasted with exact diagonalization predictions. Nevertheless, here we want
to warn the reader
that if the formation of a
bound state of 4 electrons on an otherwise empty lattice is
used as a criterion for phase separation, then the critical value near
the empty system becomes ${\rm J/t = 4.85}$ as claimed by H. Q.
Lin,\refto{lin2}
instead of a number slightly smaller than ${\rm 4}$ as shown in Fig.7. In
addition,
using exact diagonalization
methods we have evaluated the energy of the ground state at a fixed and low
electronic density,
and from there calculated numerically the second derivative with respect
to the coupling ${\rm J/t}$ to search for indications of a phase
transition. Our results suggest that phase separation may start at
${\rm J/t}$ as large as ${\sim 5.5}$. Then, the
error
bars at small density of the phase diagram Fig.7 may be large. More work
is
needed to obtain quantitative results, but nevertheless we believe that
the qualitative features of the phase diagram are properly captured by
our prediction Fig.7.

\vskip 1.9cm

\centerline{\bf III. Study of the ${\rm t-U-V}$ model}

\vskip 0.5cm

To continue our study of electronic models and the presence of
superconductivity near phase separation, let us consider the ${\rm t-U-V}$
model on a square lattice. This model is the standard one band Hubbard model
extended to include a nearest-neighbors density-density interaction
(in the study below we will analyze both an attractive and repulsive V-term).
There are several reasons to consider this model in detail. First,
it will illustrate the conjecture presented in this paper, namely the
rule that
superconducting phases of purely electronic models typically appear near
a regime of phase separation. Actually, we will discuss below that even
for the
attractive Hubbard model, which naively seems to superconduct
without phase separation in its
phase diagram, a small negative V-term is enough to induce an
instability
towards phase separation. Thus, even the attractive Hubbard model, and
thus the BCS model satisfies the rule described above in an extended
parameter space (keeping the attractive character of the potential).
In addition, we have found that one of the superconducting phases
of this model corresponds to a $\dx2y2$ condensate. The
possible existence of d-wave superconductivity in the high temperature
superconductors has received recently considerable
attention.\refto{dwave,bulut}
Then, it becomes important to have a toy model with
a condensate in this d-channel for further studies of its dynamical
properties.\refto{else}
Finally,
we believe that a possible effective model to describe holes in
the ${\rm t-J}$ model,
will include an attractive density-density interaction at distance of
one lattice spacing, and a strong repulsion on site. Such an effective
interaction is natural at least in the large ${\rm J/t}$ limit where two
holes form a tight bound state at a distance of one lattice spacing (and
of course, they cannot occupy the same site). Then, the
${\rm t-U-V}$ model in strong coupling, with ${\rm V<0}$  and ${\rm U >> 1}$
is a natural candidate for such an effective
theory.\refto{inoue} The fact that indeed it produces a $\dx2y2$ condensate
gives more
support to this conjecture.

The ${\rm t-U-V}$ model is defined as
$$
{\rm H = -t \sum_{ \langle {\bf i}{\bf j} \rangle }
( c^{\dagger}_{{\bf i}\sigma} c_{{\bf j}\sigma} +
c^{\dagger}_{{\bf j}\sigma} c_{{\bf i}\sigma} )
 + U \sum_{\bf i} ( n_{{\bf i}\uparrow} - 1/2)
( n_{{\bf i}\downarrow} - 1/2) +
V \sum_{\langle {\bf i}{\bf j} \rangle}
(n_{\bf i} - 1) (n_{\bf j} - 1 ) },
\tag {a1}
$$
\noindent where the notation is standard. It is assumed in this section
that we are working
on an $N \times N$ cluster with periodic boundary conditions.
The study will be limited to the
half-filled density, which is enough to illustrate our main results.
In this regime, the mean-field approximation used to describe the
spin-density wave (SDW) state of the repulsive Hubbard model is expected
to provide reliable information also for the ${\rm t-U-V}$
model.\refto{sdw}
Since our analysis goes beyond the Hubbard model to
include a density-density V interaction, new phases are expected to appear in
parameter space,
and thus in addition to a mean-field for the SDW state,
we will discuss a generalization for a charge-density wave (CDW), a
s-wave BCS superconducting state (SS), and a d-wave
superconducting state (DS). The appearance of SDW, CDW and SS orders are
natural since they can be obtained by perturbation theory starting from
the atomic limit (t=0) in the regimes $V=0$, $U >> 1$; $U=0$, $V >> 1$; and
$V=0$, $U<0$, $|U|>>1$, respectively. On the other hand,
the d-wave phase
is more difficult to predict intuitively.\refto{emery87}
Finally, a regime with phase separation
(PS) exists when $V<0$ and $|V| >>1$. In this region, the energy is
minimized by forming a large cluster of double occupied sites.
Of course,
numerical techniques like those presented
in the previous sections are important to verify the accuracy of the
rough mean-field predictions. For example, it is clear that the region
of phase separation will not present such a sharp division between the
region with double occupancy and the empty part of the lattice,
specially
at finite $V$.
However, the computer work for this model is
highly nontrivial, and such an analysis will be postponed
for a future publication.

Here, we will briefly describe the mean-field approach for the case of
the SDW and CDW states.\refto{sdw} For more details see also
Ref.(\cite{micnas}).
First, let us rewrite exactly the interaction
term of Eq.(\call{a1}) in momentum space as,
$$
{\rm H^{UV} = U \sum_{\bf i}
n_{{\bf i}\uparrow} n_{{\bf i}\downarrow} +
V \sum_{\langle {\bf i}{\bf j} \rangle}
n_{\bf i} n_{\bf j} = {{U}\over{4N}}
\sum_{\bf q} ( n_{\bf q} n_{\bf -q} -
\sigma^z_{\bf q} \sigma^z_{\bf -q} ) +
\sum_{\bf q} V({\bf q}) n_{\bf q} n_{\bf -q}
},
\tag {a2}
$$

\noindent where we have introduced the standard definitions of the density
and spin operators namely
$n_{\bf q} = \sum_{{\bf k} \sigma}
c^{\dagger}_{{\bf k+ q}\sigma} c_{{\bf k}\sigma}$, and
$\sigma^z_{\bf q} = \sum_{{\bf k} \sigma} \sigma
c^{\dagger}_{{\bf k+ q}\sigma} c_{{\bf k}\sigma}$. The
potential in the density-density interaction is of the form
$V( {\bf q} ) = {{V}\over{N}} ( cosq_x + cosq_y )$, and we have used
the definition $c^\dagger_{{\bf k}\sigma} = {{1}\over{\sqrt{N}}}
\sum_{\bf j} e^{i {\bf k}.{\bf j}} c^\dagger_{{\bf j}\sigma}$ in the
Fourier transformation. For the
mean-field approximation corresponding to a SDW state, we
introduce the Ansatz $\langle
n_{{\bf i}\sigma} \rangle
= {1 \over 2} [ 1 + \sigma S (-1)^{i_x + i_y} ]$ (where $S$ is a
parameter
whose value will be fixed by energy minimization, and ${\bf i} = (i_x,
i_y)$ ),
which after some straightforward algebra can be shown to be equivalent
to $\langle \sigma^z_{\bf q} \rangle = S N
\delta_{{\bf q},{\bf Q}}$, where ${\bf Q} = (\pi,\pi)$.
In addition, the constraint that
in mean value there is only one particle per site in the ground state at
half-filling,
can be formally expressed as $\langle n_{\bf q} \rangle = N \delta_{{\bf
q},0}$.
For the CDW, the  proposed Ansatz is $\langle n_{{\bf i}\sigma}
\rangle = {1\over 2} [ 1 + \rho (-1)^{i_x + i_y} ]$, which is
$\sigma$-independent, and $\rho$ is a mean-field parameter similar to $S$
for the SDW. Such an Ansatz is equivalent to requiring that
$\langle n_{\bf q} \rangle = N \delta_{{\bf q},0} + \rho N \delta_{{\bf
q},{\bf Q}}$, and $\langle \sigma^z_{\bf q} \rangle = 0$, in the mean
field ground state.

Neglecting higher order terms in the corrections to the mean-field
values, the interaction Eq.(\call{a2}) becomes,
$$\eqalign{
H^{UV}_{MF} =
&- {{U}\over{4N}} \sum_{\bf q}
( \langle n_{\bf q}        \rangle
\langle n_{\bf -q}        \rangle
- \langle \sigma^z_{\bf q} \rangle \langle
\sigma^z_{\bf -q} \rangle ) \quad + \quad
{{U}\over{2N}}
\sum_{\bf q} ( n_{\bf q} \langle n_{\bf -q} \rangle
- \sigma^z_{\bf q}  \langle \sigma^z_{\bf -q} \rangle ) \cr
&+\sum_{\bf q} V({\bf q}) ( - \langle n_{\bf q} \rangle \langle n_{\bf -q}
\rangle + 2 \langle n_{\bf -q} \rangle n_{\bf q} ).  \cr}
\tag {a3}
$$
\noindent Specializing now for the SDW state, the whole
Hamiltonian Eq.(\call{a1}) is given by,
$$
H_{MF} = \sum_{{\bf k}\sigma} \epsilon_{\bf k}
c^\dagger_{{\bf k}\sigma} c_{{\bf k}\sigma}  - \Delta \sum_{{\bf k}\sigma}
\sigma c^\dagger_{{\bf k+Q}\sigma} c_{{\bf k}\sigma} + {{U S^2
N}\over {4}}.
\tag {a4}
$$
\noindent where $\epsilon_{\bf k} = -2t ( cosk_x + cosk_y)$, the gap
$\Delta = US/2$, and $\sigma = +1$ $(-1)$ for spin up (down).
After a standard Bogoliubov diagonalization of Eq.(\call{a4}),
the ground state energy is
$$
{{E^{SDW}_{gs}}\over{N}} = - {{1}\over{N}}
\sum_{\bf k} \sqrt{ \epsilon^2_{\bf k} + \Delta^2 }
+ {{\Delta^2}  \over {U}},
\tag {a5}
$$
\noindent from which we obtain
the self-consistent equation,
$$
{1\over {2N}} \sum_{\bf k}
{{1}\over{\sqrt{\epsilon^2_{\bf k} + \Delta^2}}} = {{1}\over{U}},
\tag {a6}
$$
\noindent which has a solution for $U>0$ (note that this equation is
independent of $V$ at this level of approximation).

The procedure to obtain results for the CDW case is
very similar to that followed for the SDW state, and thus we will not
repeat it here.
The ground state energy is given by
$$
{{E^{CDW}_{gs}}\over{N}} = - {{1}\over{N}}
\sum_{\bf k} \sqrt{ \epsilon^2_{\bf k} + \Delta^2 }
- {{\Delta^2}\over{U-8V}},
\tag {a7}
$$
\noindent where we have defined $\Delta = \rho (U-8V) /2$.
The self-consistent equation for the CDW state is equal to Eq.(\call{a6}) if
$U$ is replaced by $-U+8V$. The same occurs for the energy, and thus it
is clear that the SDW and CDW states will cross at the line $U = -U +
8V$ i.e. if $U = 4V$, where both $U,V$ are positive.
Note also that the results for SDW in the V=0, positive
U axis are identical to those of the CDW state in the V=0, negative
U axis, if U is replaced by -U. This symmetry is correct even in an
exact treatment, since the repulsive Hubbard model can be exactly mapped
into the attractive Hubbard model.\refto{scalettar}

Now, let us consider the possibility of superconductivity in the phase
diagram. While it is well-known that in the $V=0$ and
negative $U$ axis, an s-wave condensate exists, other channels may become
stable when the on-site attraction is not too strong. The formalism to
handle superconductivity at the mean-field level is very standard, and
thus we will describe it only schematically. To begin with, it is
important
to exactly rewrite the interaction term defined in Eq.(\call{a2}) as,
$$
H^{UV} = \sum_{{\bf p p' q}\sigma\sigma'} \{ {{U}\over{2N}}
\delta_{\sigma -\sigma'} + {{V}\over{N}}
[ cos(p-p')_x + cos(p-p')_y ] \} c^\dagger_{{\bf p}\sigma} c^\dagger_{{\bf
-p + q}\sigma'} c_{{\bf -p'+q}\sigma'} c_{{\bf p'}\sigma}.
\tag {b2}
$$
\noindent In Eq.(\call{b2}) it is convenient to further separate the
interaction
in two pieces. One of them corresponds to ${\bf q = 0}$, and $\sigma' =
-\sigma$, which will lead to the interesting superconducting properties of
the
ground state. The other terms corresponding to ${\bf
q \neq 0}$
can be rewritten in the mean-field approximation, and after some algebra
they become
$ -{{UN}\over{4}} - 2NV + ( {{U}\over{2}} + 4V ) {\hat N}$,
where ${\hat N}$ is the total number operator (in order
to implement the constraint of working at
half-filling we have explicitly used $\langle n_{\bf q} \rangle  = N
\delta_{{\bf q}0}$).
Remember that once the Hamiltonian is written in a particle-hole symmetric form
as in
Eq.(\call{a1}),
there is no need to introduce a chemical potential to implement the
constraint of working at half-filling.
Adding together the potential and kinetic energy of the ${\rm t-U-V}$
Hamiltonian we arrive to ,
$$
H \approx  \sum_{{\bf k}\sigma} \epsilon_{{\bf k}}
c^\dagger_{{\bf k}\sigma} c_{{\bf k}\sigma}
+ \sum_{{\bf p p'}\sigma} c^\dagger_{{\bf p}\sigma} c^\dagger_{{\bf
-p}-\sigma} c_{{\bf -p'}-\sigma} c_{{\bf p'}\sigma}  f({\bf p - p'}),
 \tag {b3}
$$
\noindent where $f({\bf p - p'}) = {{U}\over{2N}} + {{V}\over{N}} [ cos(p-p')_x
+
cos(p-p')_y ]$.
To proceed with the mean field treatment, we make the standard Ansatz
$\langle c^\dagger_{{\bf p}\uparrow} c^\dagger_{{\bf -p}\downarrow}
\rangle = \langle c_{{\bf -p}\downarrow} c_{{\bf p}\uparrow} \rangle =
\phi_{\bf p}$, where $\phi$ is a real function.
The superconducting order parameter is introduced as,
$$
\Delta_{\bf p} = - 2 \sum_{\bf k} f({\bf p} - {\bf k}) \phi_{\bf k}.
\tag {c1}
$$
\noindent $\Delta_{\bf p}$ is also real, although a generalization to a complex
order parameter is very simple. With these assumptions,
the Hamiltonian Eq.(\call{b3}) can be written in the mean field approximation
as,
$$
H_{MF} = \sum_{{\bf k}\sigma}  \epsilon_{\bf k}
c^\dagger_{{\bf k}\sigma} c_{{\bf k}\sigma}
- {{1}\over{2}} \sum_{{\bf k}\sigma} \sigma \Delta_{\bf k}
( c_{{\bf -k}-\sigma} c_{{\bf k}\sigma} +
c^\dagger_{{\bf k}\sigma} c^\dagger_{{\bf -k}-\sigma} ) +
\sum_{\bf k} \Delta_{\bf k} \phi_{\bf k}.
\tag{c2}
$$
\noindent The diagonalization of Eq.(\call{c2}) is performed
following standard techniques, and thus we simply present the results.
The ground state energy is,
$$
{{E_{gs}}\over{N}}
= \sum_{\bf k} \Delta_{\bf k} \phi_{\bf k}
- {{1}\over{N}} \sum_{\bf k} \sqrt{ \epsilon^2_{\bf k} + \Delta^2_{\bf k} },
\tag{c3}
$$
\noindent and the self-consistent equation obtained from minimization of the
ground state energy is
$$
\Delta_{\bf p} = - \sum_{\bf k} {{\Delta_{\bf k}
f({\bf k} - {\bf p}) } \over
{ \sqrt{ \epsilon^2_{\bf k} + \Delta^2_{\bf k} } } }.
\tag{c4}
$$
\noindent
Now, let us specialize the generic results Eq.(\call{c3}) and
(\call{c4}) to the particular case of an isotropic s-wave condensate (SS). This
symmetry is certainly relevant for the attractive Hubbard model
($V=0$, $U<0$). For an s-wave we can assume $\Delta_{\bf p}$ to be
${\bf p}$-independent, and the self-consistent equation formally becomes again
Eq.(\call{a6}), replacing $U$ by $-U$. The s-wave ground state energy is
$$
{{E^{SS}_{gs} }\over{N}} =
{{\Delta^2}\over{|U|}}
- {{1}\over{N}} \sum_{\bf k} \sqrt{ \epsilon^2_{\bf k} + \Delta^2 }.
\tag{c5}
$$
\noindent Note that this energy, and thus the self-consistent equation
derived from it, are independent of $V$, as in the case of the SDW
state. As explained before, it is also interesting to notice that for the
purely attractive
Hubbard model ($V=0$, $U<0$) the energy of the CDW state and the
superconducting SS state are degenerate. This is to be expected since by
a simple transformation the attractive and repulsive Hubbard models can
be mapped into each other. The CDW (SS) correlations become the spin
correlations in the Z (XY) direction. By rotational invariance, the Z
and XY spin correlations are identical in the repulsive Hubbard model,
implying the degeneracy of the CDW
and SS states for the attractive case, even in an exact treatment of the
problem.

The gap equation Eq.(\call{c4}) also admits solutions in channels other
than s-wave. Let us consider the Ansatz $\Delta_{\bf k} =
\Delta_0 ( cosk_x - cosk_y)$ which corresponds to a $\dx2y2$
superconductor
($\Delta_0$ is assumed to be momentum independent, and will be obtained
by self-consistency). After considerable but straightforward algebra, we
arrive to the self-consistent equation for $\Delta_0$,
$$
{{1}\over{|V|}} = {{1}\over{N}} \sum_{\bf k} { {cosk_x ( cosk_x - cosk_y)}
\over
{\sqrt{
\epsilon^2_{\bf k} + \Delta^2_0 ( cosk_x - cosk_y)^2 } } }.
\tag {c9}
$$
\noindent This equation can be solved iteratively for $\Delta_0$.
The ground state energy for this condensate is,
$$
{{E^{DS}_{gs} }\over{N}} = {{\Delta^2_0}\over{|V|}}
-{{1}\over{N}} \sum_{\bf k} \sqrt{ \epsilon^2_{\bf k} +
\Delta^2_0(cosk_x - cosk_y)^2 }.
\tag{c8}
$$
\noindent Whether this d-wave state becomes stable or not in some
region of parameter space, depends on its energy competition with the
other possible states.
Finally,
in the region $V <0$ (and for large $|V|$), there is phase separation
i.e. it is energetically preferable to separate the lattice into a
region mostly
without electrons, and a region where every site is doubly occupied.
In this way, the dominant $V$-term of that regime is minimized. Including
fluctuations, the separation between the two regions will not be sharp
and it may occur that the two dominant states have densities $\langle n
\rangle = 1 + \delta$ and $1 - \delta$, with the parameter
$\delta$ different from 1 for a finite value of $|V|$. Preliminary
Quantum Monte Carlo results support this scenario.
However, for our
rough mean-field approximation, we will assume a ``perfect'' phase
separation with $\delta = 1$ (i.e. where electrons are simply not
allowed into the empty part of the lattice),
and thus the ground state energy is simply,
$$
{{E^{PS}}\over{N}} = {U \over 4} + 2V.
\tag {b1}
$$

The phase diagram of the ${\rm t-U-V}$ model in the mean field approximation is
obtained by comparing the energies of the SDW, CDW, SS, DS, and PS
phases given by Eqs.(\call{a5}),(\call{a7}),(\call{c5}),(\call{c8}),
and (\call{b1}), respectively. The results are presented in
Fig.8, showing that the phase diagram at half-filling is very rich. In the
purely
repulsive Hubbard regime and its neighborhood, the SDW state has the
lowest energy, as expected. Increasing $V$ a transition to the CDW regime
is found at $U = 4V$, as was explained before.
This CDW phase is very robust and is stable also in the presence of a negative
$U$ term. Actually, in the purely attractive Hubbard model axis, the
CDW state is degenerate with the superconducting state SS.
If a small and negative $V$ term is introduced, then
this SS state becomes stable, and the CDW-SS degeneracy is broken.
However, decreasing
further the strength of $V$ (towards more negative values), induces another
transition into
a phase separated regime. Thus, in this model we also observe the feature
discussed in the
introduction namely that
a superconducting
phase appears in the neighborhood of phase separation, as it occurs in
the 2D ${\rm t-J}$ model. In this case it is necessary to extend the
parameter space beyond an on-site attractive
${\rm U/t}$ term, to include further attractive interactions at a distance of
one lattice spacing in order
to observe this behavior. Such an extension is natural, since a
strictly on-site force is somewhat pathological and
physically difficult to realize.\refto{comen}
Phase separation is very robust in this model as
shown in Fig.8, and it exists even for large and positive values of
${\rm U/t}$.
Finally, it is interesting to observe the presence of an ``island'' in
parameter space where the $\dx2y2$ state is stable. This occurs for
small
values of $|U|$, and negative $V$. Again, increasing further $|V|$ leads to an
instability towards phase separation. Note that for large values of
${\rm U/t}$, there is a direct transition from SDW to phase separation.
Then, finding phase separation in a given model is not sufficient to
guarantee the
presence of superconductivity in its vicinity. More details about the
interesting phase diagram shown in Fig.8, specially regarding the d-wave
condensate,
will be presented in a future publication.

\vskip 1.9cm

\centerline{\bf IV. Two band Hubbard model on a chain}

\vskip 0.5cm

The presence of superconductivity near phase separation is not
restricted to one band models. The same phenomenon occurs in
the ground state of multi-band Hamiltonians, and in this section
we analyze a particular (and physically relevant) case in detail.
The model we will study is the
three band model proposed by Emery,\refto{emery87} and Varma and
collaborators,\refto{varma}
which is believed to contain the basic ingredients to
describe the behavior of electrons in the ${\rm Cu O_2}$ planes of the
high-Tc superconductors.
Using the $hole$ notation,
where the vacuum is defined as having all the $d$-orbitals occupied in the
copper sites and the $p$-orbitals occupied in the oxygen sites, the
Hamiltonian is
$$\eqalign{
{\rm H =}
&{\rm -t_{pd}\sum_{\bf \langle i j \rangle} p^\dagger_{\bf j}( d_{\bf
i} + h.c.) + \epsilon_d \sum_{\bf i}  n^d_{\bf i}
+ \epsilon_p \sum_{\bf j}  n^p_{\bf j} } \cr
&+ {\rm U_d \sum_{\bf i} n^d_{{\bf i}\uparrow} n^d_{{\bf i}\downarrow} +
U_p \sum_{\bf j} n^p_{{\bf j}\uparrow} n^p_{{\bf j}\downarrow} +
V \sum_{\bf \langle i j \rangle}
n^d_{{\bf i}} n^p_{{\bf j}} },
\cr}
\tag {gg}
$$
\noindent where ${\rm p_{\bf j}}$ are fermionic operators that destroy
holes
at the oxygen sites labeled ${\rm {\bf j}}$,
while ${\rm d_{\bf i}}$ corresponds to hole annihilation
operators at the copper sites ${\rm {\bf i}}$. ${\rm {\bf \langle i j
\rangle} }$ refers to pairs of nearest neighbors ${\rm {\bf i}}$ (copper) and
${\rm {\bf j}}$ (oxygen) sites
(and ${\rm t_{pd}}$ is the corresponding hybridization between
copper and oxygen).
${\rm U_d}$ and ${\rm U_p}$ are positive constants
that represent the repulsion between
holes when they are located at the same d and p orbitals, respectively. ${\rm
V}$
corresponds to the
Coulombic repulsion when two holes occupy adjacent ${\rm Cu-O}$ links.
The importance of this term has been remarked by Varma and
collaborators.\refto{varma}
In principle, interactions at larger
distances should also be included in the Hamiltonian, but such an
analysis will not be carried out in this paper.\refto{future}
$\epsilon_d$ and $\epsilon_p$ are the energies of each orbital (with the
charge transfer constant defined as $\Delta = \epsilon_p - \epsilon_d$,
which for the cuprates is a positive number). The doping fraction is
defined
as ${\rm x = n_h/N}$, where ${\rm n_h = (N_h - N)}$ is the number of
holes doped away from half-filling, ${\rm N_h}$ is the total number of holes,
and ${\rm N}$ is the
number
of ${\rm Cu-O}$ cells.
At half-filling ${\rm N_h = N}$, as in the insulating parent cuprates.
Below, only the region
${\rm 0 \leq x \leq 1}$ will be explored which is relevant for
hole-doped materials.

It is generally accepted that ${\rm U_d}$ is the largest parameter of
the model. In particular ${\rm U_d} > \Delta$, which is consistent with
the
charge transfer character of the copper oxides. It is also generally
accepted
that in the undoped case there is one hole in each ${\rm Cu}$ site,
while upon doping, additional holes are introduced into the oxygen
sites.
Under these assumptions (supplemented with ${\rm V=0}$), Zhang and
Rice\refto{singlet}
derived the ${\rm t-J}$ model as an effective low energy Hamiltonian of
the
more general three band model. In the region of parameter space where
this
derivation is valid, the conclusions we reached for the ${\rm t-J}$
model
in the previous sections will also apply to the more general Emery-Varma
Hamiltonian Eq.(\call{gg}). However, the three band model is more
general than the ${\rm t-J}$ model and thus it is worth exploring the
behavior of its ground state in regions of parameter space that cannot
be mapped into a simplified one band
Hamiltonian. For example, in the particular case
${\rm U_d = \infty}$, an interesting region of phase separation was studied
with the slave-boson approach
for finite ${\rm V}$, and superconducting instabilities
were observed in its vicinity.\refto{grilli}
Then, phenomena similar to those described in sections II and III also
take place in the multiband Hubbard model.

Recently, a more detailed study of superconductivity near phase
separation
in the Emery-Varma model,
was discussed using exact diagonalization
techniques\refto{sudbo2,sudbo,sano}
applied to the one dimensional version of Eq.(\call{gg}) (i.e. for the
``two band'' Hubbard model). Part of the results of these studies (which go
beyond the
mean-field approximation) are
summarized in Fig.9. In this figure, the copper (oxygen) sites
are represented by large (small) circles. The ground state at
half-filling (${\rm n_h=0}$) is shown in the atomic limit ${\rm t_{pd} = 0}$ in
Fig.9a
(where the direction of the spins is arbitrary). There is one hole per
copper site. Fig.9b shows the ground
state introducing two holes
(${\rm n_h = 2}$) and assuming ${\rm V} < \Delta + {{3}\over{4}} {\rm U_p}$. In
this case the holes populate the oxygen sites. However, if
${\rm V} > \Delta + {{3}\over{4}}
{\rm U_p}$,
the ground state drastically changes to the one
shown in Fig.9c, where in addition,
it is assumed that ${\rm U_d} > {{3}\over{2}}({\rm U_p} + 2 \Delta) $,
otherwise doped holes would prefer to be located in copper sites in the
limit ${\rm V = \infty}$. In this case, the electrons tend to occupy the
oxygens depopulating the coppers.

In Fig.9c, the minimum energy is obtained by the formation of
``biexcitons'' (in the language of Ref.(\cite{sudbo2})), and thus a charge
transfer instability
from
copper to oxygen ions occurs. This same phenomenon takes place in two
dimensions
as shown in Fig.9d also for the case of ${\rm n_h = 2}$, and large
${\rm V}$. When additional
holes are added to the system, the region of doubly occupied oxygens
increases in size. Then, the
nearest neighbor repulsion ${\rm V}$ leads to the formation of tight hole
bound states at the oxygen sites. In turn, this leads to
phase separation which occurs between a phase with a density of one particle
per
cell (with the charges on the copper sites), and a region of density of
two particles per cell (with all the charge on oxygens).
However, remember that
the results summarized in Fig.9 were obtained in the atomic limit. When
${\rm t_{pd} \neq 0}$ the
actual phase boundaries, and even the existence of phase separation, depends on
the interplay between the kinetic energy and the Coulomb terms. In order
to study this interplay, exact diagonalization techniques were recently
applied\refto{sudbo,sano} to finite chains for particular
values of the parameters of the two band model.
Indications of superconductivity were observed
in a special region of parameter space by studying
the value of the
parameter ${\rm K_{\rho}}$ used in conformal field theory, and by
the
analysis of the anomalous flux quantization in the presence of an
external
flux through the ring (i.e. closing the chain with periodic boundary
conditions).

In this section,
we also use Lanczos diagonalization techniques to obtain ground
state properties of the one dimensional version of Eq.(\call{gg}), and
analyze further the interplay between superconductivity and phase
separation, according to the ideas discussed in the introduction.
Due to the rapid growth of the Hilbert space of the problem with the
chain
size, our analysis is limited to 6 cells (12 sites), periodic
boundary
conditions, and ${\rm n_h
= 2, 4}$ and ${\rm 6}$ holes. In order to work in the regime of hole pairing
in oxygen sites, we select ${\rm U_d = 7}$, ${\rm U_p = 0}$, and $\Delta
= 1.5$, all in units of the hopping integral ${\rm t_{pd}}$. The
parameters satisfy the
relation ${\rm U_d > {{3}\over{2}} ( U_p + 2 \Delta ) }$, and are
similar to those used by Sudbo et al.\refto{sudbo} The parameter ${\rm
V}$ is varied between 0 and 8, which is large enough to reach the phase
separated regime. A quantitative indication of the charge transfer from
${\rm Cu}$ to ${\rm O}$ sites, and eventually of phase separation, is
given
by the average occupation of oxygen sites. Alternatively, a more clear
indicator of the crossover from the state shown in Fig.9b to Fig.9c is
given by the susceptibility $\chi_{ps} = \langle n_O^2 \rangle - \langle n_O
\rangle^2$ (see Ref.(\cite{scale})). This quantity is maximized when a transfer
of
charge from copper to oxygen takes place, i.e. at the phase separation
transition,
and then it slowly decays to zero at large ${\rm V}$ once phase
separation has already occurred, since the
fluctuations
in the occupation number of the oxygens vanish in this limit.
An independent determination of the crossover to the phase separated
regime
is given by the short wavelength component of the susceptibility
associated
with the correlations of pairs of holes in the oxygen
sites.\refto{assaad}
This quantity (that we called X) is normalized such that it is equal to 1 for
${\rm V
\rightarrow \infty}$ if the phase separated state Fig.9c is reached in this
limit. For details see Ref.(\cite{assaad}).

The numerical results for the  susceptibility $\chi_{ps}$ are shown in Fig.10a
for the case of two holes ${\rm n_h = 2}$, and hopping
${\rm t_{pd} = 0.5}$, and ${\rm t_{pd} = 1}$.
In the atomic limit (${\rm t_{pd} = 0 }$),
the crossover between the two states
is sharp, and takes place at ${\rm V} = \Delta = 1.5$ (not shown in the
figure). For a fixed $\Delta$, the value of ${\rm V}$ at which the
crossover occurs increases with ${\rm t_{pd}}$, and it is located
between ${\rm 2 \leq V \leq 3}$ for ${\rm t_{pd} = 1}$. The order
parameter X has its maximum variation also in the same interval, making
the results compatible among themselves and with the simple picture
shown
in Fig.9a-d. On the other hand, the results corresponding to ${\rm n_h =
4}$ are very different. The order parameter X does not saturate to 1 at
large ${\rm V}$, and the occupation number of oxygen sites is maximized
in this limit (i.e. apparently there is no occupancy of the copper
sites!). However, this result is a finite size effect.
To visualize this
problem, simply add two more holes to the 6 cells chain shown in Fig.9c.
In the atomic limit the possible states are 5 double occupied oxygens
or 4 double occupied oxygens and 2 single occupied oxygens, and thus
strictly speaking the phase separated regime defined before cannot be
realized due to the particular size of the chain, and the 5 pairs of
holes have a finite mobility (as in a doped attractive Hubbard chain).
This effect does not occur on a larger chain (as example consider 20 holes
on a 12 cells chain which keeps the density constant).
Then, the results of Fig.10b should not be considered representative of
the bulk limit, and they are shown here mainly as a warning to the
reader
that finite size effects have to be carefully controlled in these
numerical studies.

In order to study the existence of a superconducting phase arising from
the mechanism of hole pairing at the oxygen sites, we studied the
pairing correlation,
$$
C(m) = {{1}\over{N}} \sum_{j} \langle \Delta^\dagger_{j+m} \Delta_j \rangle,
\tag {gg2}
$$
\noindent where the pairing operator is defined as $\Delta_j =
c_{j\uparrow} c_{j\downarrow}$, and $j$ denotes oxygen sites. The
results
for the pairing correlations corresponding to the chain of 6 cells, and
the same values of parameters as used in Fig.10a, are shown in Fig.11a
for ${\rm n_h = 2}$, at several values of ${\rm V}$.
It can be seen that the pairing correlation at the largest
distance slowly
increases with ${\rm V}$ up to ${\rm V = 4}$ (measured with large steps
in ${\rm V}$ of ${\rm 2}$ in units of ${\rm t_{pd}}$),
i.e. approximately
in the crossover to the
phase separated regime. At ${\rm V=6}$, phase separation has been
reached and the double occupied oxygen sites are in contiguous
sites forming a rigid structure with very low mobility. Consequently,
the pairing correlations are suppressed at large ${\rm V}$.
Then, Fig.11a shows a behavior
qualitatively very similar to those reported before for the ${\rm t-J}$
model and the ${\rm t-U-V}$ model i.e. the region where superconducting
correlations exist is the vicinity of phase separation. However, note
that
the size of the tail in the pairing correlation is much smaller than
that
observed for the ${\rm t-J}$ model in one and two dimensions.\refto{super,prl}
Actually, we measured the pairing correlations at the same
parameters used before by Sudbo et al.,\refto{sudbo} and also in the
superconducting
region studied by Sano and
Ono\refto{sano} in the ${\rm U_d = \infty}$ limit.
We observed that the tail in
$C(m)$ is negligible in both cases, and thus only the conformal field theory
parameter
$K_{\rho}$ is left for the analysis of pairing in this region (it is
expected that $K_{\rho}$ will present smaller finite size effects that
the actual correlations). This is unfortunate since to study other
properties of the superconducting condensate (specially dynamical
properties)
it is important to have robust pairing correlations developed in the
ground state of the clusters that can be studied numerically.
Note also that the superfluid density does not
exist in one dimension since for its definition a careful two
dimensional limit in the current correlations needs to be
considered,\refto{ds} as mentioned briefly in Appendix A. One should
be careful in not confusing the Drude weight and the superfluid density
which are obtained from very similar current correlations.

In Fig.11b, the results corresponding to ${\rm n_h = 4}$ are shown,
again
for several values of ${\rm V}$. The pairing correlations are clearly
more robust than those shown at smaller hole doping in Fig.11a. However,
this effect is unfortunately
spurious and caused by the finite size of the chain as discussed before.
The 10 holes prefer to be located on oxygen sites, with none on copper
sites, and thus the oxygen pairs have enough mobility to induce a robust
pairing signal.
We are currently investigating the possibility of enlarging the range of
the repulsive density-density interaction in order to stabilize the
results of Fig.11b. Work is in progress.
Finally, in the case ${\rm n_h = 6}$ all oxygen sites are doubly
occupied in the large ${\rm V}$ limit, and we observed that $C(m)$
decays to zero rapidly with distance as expected.

In short, a numerical analysis of the two band Hubbard model on a chain
roughly shows a behavior similar to that of the ${\rm t-J}$ model in 1D
and 2D, and of the ${\rm t-U-V}$ model namely the presence of superconducing
correlations near half-filling. A study of the same phenomenon for a
three band model in 2D would be important.

\vskip 1.9cm

\centerline{\bf V. Conclusions}

\vskip 0.5cm

In this paper we have analyzed several models of correlated electrons
using
both numerical and analytical techniques. In particular, we search for
superconducting phases as a function of couplings and densities. Our
main result is that the two dimensional ${\rm t-J}$ model has a region
of $\dx2y2$ superconductitivy near phase separation centered at a density of
quarter-filling $\langle n \rangle = 1/2$. In addition, an interesting
transition
from d-wave to s-wave superconductivity was observed reducing the
electronic fermionic density. Although the presence of the d-wave
superconducting
phase for realistic densities remains unclear, we
argued that numerical studies in this region may be affected by
the small number of pairs contributing to the signal. Then, we believe
that the possibility of finding superconductivity in this model in the
realistic region of small ${\rm J/t}$ and densities close to
half-filling
is still open. The most favorable channel in this region is clearly the
$\dx2y2$. Now that superconductivity has been identified in the phase
diagram,
we believe that the most suitable procedure to
follow
is to describe the quarter filling region with a variational wave
function of condensed d-wave pairs, such that a good agreement with
the numerical work is found, and then the variational calculation should
be repeated at other densities on larger clusters.

In addition, in this paper we conjectured that the presence of
superconductivity near
phase separation observed for the ${\rm t-J}$ model may be a general feature of
several models of correlated electrons. To explore this possibility we
studied analytically the ${\rm t-U-V}$ model in two dimensions at
half-filling. Indeed we observed superconductivity near phase
separation in the regime of attractive couplings.
For large and negative ${\rm U}$ the condensate is s-wave,
while increasing ${\rm U}$ it becomes $\dx2y2$, as for the ${\rm t-J}$
model. We also observed signals of superconductivity near phase
separation for the one dimensional two band Hubbard model. In this case
the analysis was done using numerical methods.
As a thumb-rule we believe that once an electronic model is proposed to
describe a superconducting material, first it is convenient to search for
indications of phase separation in the phase diagram, which is usually
not very difficult, and then superconductivity should be analyzed in its
boundary. This rule seems to work in all models of correlated electrons
that we are aware of (at least those with finite range interactions).


\vskip 1.cm

\centerline{\bf Acknowledgments}

\vskip 0.5cm

We thank D. Scalapino, S. Maekawa,
A. Kampf, T. K. Lee, V. Emery and G. Zimanyi for useful comments.
E. D. and A. M. are supported by the Office of Naval Research under
grant
ONR N00014-93-1-0495.
J. R. has been supported in part by the U. S. Department of Energy (DOE)
Office of Scientific Computing under the High Performance Computing and
Communications Program (HPCC), and in part by DOE under contract No.
DE-AC05-84OR21400 managed by Martin Marietta Energy Systems, Inc., and
under
contract No. DE-FG05-87ER40376 with Vanderbilt University.
Y. C. C. is supported by the National Science Council of R. O. C. under
the Grant NSC82-0511-M007-140.
We thank the
Supercomputer Computations Research Institute (SCRI) for its support.
Part of the computer calculations were also done at the CRAY-2 of the
National Center for Supercomputing Applications, Urbana, Illinois.
\endpage

\centerline{\bf Appendix A: study of the superfluid density}

\vskip 0.5cm

For completeness, in this appendix we remind the reader of an
alternative
way to characterize a superconducting phase. In the bulk of this paper, we have
used the existence of long-range order
in the equal-time pairing correlation as an indication of superconductivity.
However,
there is another approach that does not involve the study of
different symmetry sectors. This technique consists in the
evaluation of the ``superfluid density'' ${\rm {\ds}}$ in the region
of interest. It has been recently shown\refto{ds} that this
quantity can be obtained on a finite cluster following steps similar
to those necessary to calculate
the Drude weight.\refto{dago92} Actually, it can be shown that
$$
\quad\quad\quad\quad\quad\quad\quad\quad\quad
{\rm {{\ds}\over{2 \pi e^2 }} =
{{\langle -T \rangle} \over{4N}}  - {{1}\over{N}}
\sum_{n \neq 0} {1\over{E_n - E_0}}
|\langle n | j_x({\bf q}) |0\rangle |^2 },
\quad\quad\quad\quad\quad\quad\quad\quad (A.1)
$$
\noindent where ${\rm e}$ is the electric charge;
the current operator in the x-direction with momentum ${\rm {\bf q}}$
is given by ${\rm j_x ({\bf q})} = \sum_{{\bf l},\sigma}
e^{i {\bf q}.{\bf l}}
( {\bar c}^{\dagger}_{{\bf l},{\sigma}}
  {\bar c}_{{\bf l + {\hat x}},{\sigma}} -
  {\bar c}^{\dagger}_{{\bf l}+{\bf {\hat x}},{\sigma}}
  {\bar c}_{{\bf l},{\sigma}} )$; $\langle {\rm - T} \rangle$ is the
mean value of the
kinetic energy operator of the model under study; ${\rm | n \rangle}$ are
eigenstates
of the Hamiltonian with energy ${\rm E_n}$ (where ${\rm n = 0}$ corresponds
to the ground state), and the rest of the notation is standard.
The momentum ${\rm {\bf q }=(q_x,q_y)}$ of the current operator
needs to be selected such that
${\rm q_x}=0$ and ${\rm q_y \rightarrow 0}$.
The constraint of having
a small but nonzero ${\rm q_y}$ is necessary to
avoid a trivial cancellation of $\ds$
due to rotational and gauge
invariance.\refto{ds}
On the $4 \times 4$ cluster, the minimum value of ${\rm q_y}$
is $\pi/2$, which unfortunately is not small.
$\ds$ given by Eq.(A.1) can be
evaluated numerically using a continued fraction expansion
technique.\refto{dago92}

Results for ${\rm {\ds}}$ have already been presented for the
2D ${\rm t-J}$ model close to phase separation in Ref.(\cite{prl}).
The superfluid density has a clear peak in the same region where the
pairing correlations are maximized. Thus, the results obtained with
these quantities are consistent with each other, and they support the
conclusion that a superconducting phase exist in the 2D ${\rm t-J}$ model.
Note also that in Ref.(\cite{prl})
it was shown that the $Drude$ peak is very large not only
in the superconducting phase, but also for smaller values of ${\rm J/t}$.
This is to be expected since a small resistivity (actually zero in the bulk
limit) does not uniquely mean that the system superconducts, since a perfect
metal
has also zero resistivity. For details see Ref.(\cite{prl}),
Ref.(\cite{ds}),
and Ref.(\cite{review}).

It is interesting to remark that we have carried out a study of ${\rm
{\ds}}$ in the one band Hubbard model with a repulsive interaction on a
$4 \times 4$ cluster.\refto{fano} Unfortunately, we have not observed
any indication of superfluidity in this model based on ${\rm {\ds}}$.
Since other groups may attempt such a study, it would be helpful for
them to have some numerical results to compare with, and thus we provide them
here: working at ${\rm U/t} = 8$ (which is the most realistic region of
parameter space for
the one band Hubbard model as a model of high Tc cuprates),
and at the representative  density of 14 (8) electrons, we found
that ${\rm {{\langle -T \rangle}\over{4N}} = 0.26473 }$ ($0.28028$). The
complicated second term of the r.h.s. of Eq.(A.1) evaluated at momentum
${\bf q} = (0,\pi/2)$
gives $-0.33636$ and $-0.34836$, for 14 and 8 electrons, respectively.
Then, ${\rm D_s/ (2 \pi e^2) } = -0.07163$, and $-0.06808$, again
for 14 and 8 electrons, respectively. Note that
a negative result for ${\rm D_s}$ is not impossible on
a finite system, and similar problems have been observed in the analysis
of the Drude peak near half-filling for the one band Hubbard model.\refto{fye}
Also
in the ${\rm t-J}$ model, at $small$ ${\rm J/t}$, the superfluid density
is negative
on finite clusters. These results should be taken as indicative that
there
are $no$ superconducting correlations in these regimes of parameter space.

\vskip 1.0cm

\centerline{\bf Appendix B: hard-core bosons in the t-J model}

\vskip 0.5cm

In section II the phase diagram of the two dimensional ${\rm t-J}$ model
was analyzed, finding indications of superconductivity near phase
separation. This result is qualitatively similar to that found in the one
dimensional version of the same
model.\refto{ogata} It is interesting to notice that
in one dimension the $statistics$ of the particles described by the
${\rm t-J}$ model is irrelevant since a pair of them cannot be interchanged
due to the constraint of no double occupancy at every site.\refto{com3}
In other words, fermions and hard-core bosons produce the same physics
in this model. Other one dimensional models have the same property.
Then, a natural question arises:
is the role of the statistics in two dimensions important
for the qualitative features of the phase diagram? In order to
study this problem we analyze the same ${\rm t-J}$ model defined in
section II but removing the fermionic
statistics i.e. considering hard core bosons
with spin-1/2. This is an artificial model without physical realization
(to the best of our knowledge) but mathematically well-defined, and
its analysis will teach us whether the signs coming from fermionic
permutations are important for the phase diagram.\refto{long} As an additional
motivation note that sometimes the treatment
of hard-core bosons is simpler than
that of fermions. In particular, it may occur that Monte Carlo
simulations without ``sign-problems'' can be carried out for bosons and not
for fermions. Then, it is important to know how drastic an approximation
would be to neglect the statistics in the two dimensional ${\rm t-J}$ model.

Using a $4 \times 4$ cluster and exact diagonalization techniques,
we investigated the quantum numbers of the
ground state, and the sign of
$\Delta_B = E(n+2) + E(n) - 2 E(n+1)$ as a criterium to analyze the
presence
of binding of particles in the system (where $E(m)$ is the ground state energy
in the
subspace of $m$ holes). In an analogous way, we search for indications
of phase
separation studying the sign of $\Delta_{PS} = E(n+4) + E(n) - 2
E(n+2)$. Although the actual results are somewhat erratic (probably due
to finite size effects), the analysis of these numbers show a clear
pattern which is schematically shown in Fig.12. The region of phase
separation is robust and not appreciably affected by the statistics of
the particles. However, the metallic region of the phase diagram
drastically changes replacing fermions for hard-core bosons. A large
``ferromagnetic'' region is observed at all densities. The total spin
is maximized (fully polarized ferromagnet) for small ${\rm J/t}$, and
then
it decreases until the boundaries of the phase are reached.
It can be shown that even two particles
in an otherwise empty lattice minimize their energy by forming a S=1
state, at least for ${\rm J/t}$ smaller than some finite number and on a
finite cluster.
Numerically we systematically observed a small window between
the ferromagnetic and phase separated regimes. In this regime, we found
$\Delta_B
< 0$ suggesting the presence of pairing. However,
it is not clear whether
this small detail will survive the bulk limit increasing the lattice size.
Then, we conclude that the phase diagram of the two dimensional ${\rm
t-J}$ model is strongly affected by the statistics of the particles, and
approximations that do not handle this properly may produce incorrect results.

\endpage

\singlespace

\references

\singlespace

\refis{ohta} S. Maekawa, private communication; Y. Ohta, T. Shimozato,
R. Eder, K. Tsutsui, and S. Maekawa, preprint in preparation.

\refis{comen} However, it is fair to point out that in the case of the
${\rm t-U-V}$ model the attractive force that produces pairing (U-term) is not
strictly the same that induces phase separation (V-term). However, we
can think of them as forming part of a unique attractive
force that decays with distance
by moving in parameter space along straight lines ${\rm U \propto V}$ in
the quadrant where both ${\rm U}$ and ${\rm V}$ are negative.

\refis{trunca} A possible algorithm to use for the study of large
clusters is the ``many-body basis set
reduction'' discussed in
J. Riera and E. Dagotto, \phys {\bf 47}, 15346 (1993).

\refis{rpa} Very recently we received a paper by D. van der Marel
(Groningen preprint, Aug. 1993) where the RPA approximation is used
to find the phase diagram of the ${\rm t-J}$ model. In excellent
agreement with our results, a region of d-wave superconductivity
was found near half-filling, with a transition to extended s-wave
superconductivity
at low electronic doping.

\refis{micnas} R. Micnas, J. Ranninger, and S. Robaszkiewicz, Rev. Mod.
Phys. {\bf 62}, 114 (1990).

\refis{tklee} Y. C. Chen and T. K. Lee, Aug. 1993 preprint, also studied the
low density regime of the ${\rm t-J}$ model in two dimensions finding a region
of s-wave pairing compatible with our results.

\refis{future} A study of the two bands Hubbard model including long-range
interactions
is in preparation by two of the authors (J.R. and E.D.).

\refis{com3} In principle, working on a finite
ring the statistics of the particles can
be noticed through hopping processes that use the
boundary. But in the bulk limit of the ${\rm t-J}$ model on a chain, the
statistics should not affect the results.

\refis{fye} R. Fye et al., \phys {\bf 44}, 6909 (1991).

\refis{chen} Y. C. Chen, and T. K. Lee, \phys {\bf 47}, 11548 (1993).

\refis{inoue} J. Inoue, and S. Maekawa, Prog. of Theor. Phys., Suppl.
{\bf 108}, 313 (1992).

\refis{else} A more complete study of this model including corrections
to the mean-field approximation, and an extensive numerical analysis of
its properties is in preparation, and will be presented in a future
publication.

\refis{foot} Note, however, that this
cluster is not invariant under reflections.

\refis{gros} C. Gros, \prb 38, 931, 1988;
see also C. Gros, R. Joynt, and T. M. Rice, Z. Phys. {\bf B 68},
425 (1987) and
G. J. Chen, R. Joynt,
F. C. Zhang, and C. Gros, \phys {\bf 42}, 2662 (1990).

\refis{valenti} R. Valenti and C. Gros, \prl 68, 2402, 1992.

\refis{lin2} H. Q. Lin, \prb 44, 4674, 1991.


\refis{scale} R. T. Scalettar, D. J. Scalapino, R. L. Sugar, and S. R.
White, \phys {\bf 44}, 770 (1991).

\refis{assaad} F. F. Assaad and D. Wurtz, \phys {\bf 44}, 2681 (1991).

\refis{oitmaa} The ``square'' lattices that can cover the bulk, and
also hold a perfect N\'eel order with periodic boundary conditions
are those satisfying ${\rm N = n^2 + m^2}$, where ${\rm N}$ is the
number of sites, and ${\rm n},{\rm m}$ are integers, with ${\rm n +m}$
even (see J. Oitmaa and D. Betts, Can. J. Phys. {\bf 56}, 897 (1978)).

\refis{review} E. Dagotto, ``Correlated Electrons in High Temperature
Superconductors'', NHMFL preprint (1993) (to appear in Rev. Mod. Physics).

\refis{muller} J. Bednorz and K. M\"uller, Z. Phys. {\bf B 64}, 188
(1986); Rev. Mod. Phys. {\bf 60}, 585 (1988).

\refis{putikka} W. Putikka, M. Luchini and T. M. Rice, \prl 68, 538, 1992.

\refis{prl} E. Dagotto and J. Riera, \prl 70, 682, 1993.

\refis{barnes} T. Barnes, Int. J. of Mod. Phys. {\bf C2}, 659 (1991).

\refis{dcj}
D. C. Johnston, Phys. Rev. Lett. {\bf 62}, 957 (1989).

\refis{schilling}
A. Schilling, et al., Nature {\bf 363}, 56 (1993);
S. N. Putilin, E. V. Antipov, O. Chmaissem, and M. Marezio, Nature {\bf
362}, 226 (1993).

\refis{dago92} E. Dagotto, A. Moreo, F. Ortolani, D. Poilblanc, and
J. Riera, \phys {\bf 45}, 10741 (1992).

\refis{batlogg}
B. Batlogg, Physics Today June, page 44 (1991);
B. Batlogg, H. Takagi, H. L. Kao, and J. Kwo,
``Electronic properties of High Tc Superconductors, The Normal and
the Superconducting State.'', Eds. Kuzmany et al., Springer-Verlag (1992).

\refis{allen}
R. O. Anderson, et al., \lett {\bf 70}, 3163 (1993).

\refis{foot2}
A calculation along
these lines has been reported by Valenti and Gros.\refto{valenti} The position
of the transition to phase separation at quarter filling is in excellent
agreement with the numerical results (i.e. ${\rm J/t \sim 3}$, see their
Fig.1 including the density-density attraction in the Hamiltonian).
However, their wave function does not show indications of superconductivity at
this doping (only at low electronic densities). Then, the Valenti-Gros
wave function does not seem accurate enough to describe the physics of the
${\rm
t-J}$ model at quarter-filling.

\refis{bulut}
N. Bulut, and D. Scalapino, \lett {\bf 67}, 2898 (1991);
P. Monthoux, A. Balatsky, and D. Pines, \lett {\bf 67},
3448 (1991).

\refis{cheong}
S.-W. Cheong, et al., \lett {\bf 67}, 1791 (1991).

\refis{fukuyama}
T. Tanamoto, H. Kohno, and H. Fukuyama, J. Phys. Soc. Jpn. {\bf
61}, 1886 (1992); J. Phys. Soc. Jpn. {\bf 62}, 717 (1993);
J. Phys. Soc. Jpn. {\bf 62}, 1455 (1993).

\refis{prl91}
E. Dagotto, A. Moreo, F. Ortolani, J. Riera, and D. J. Scalapino, \lett
{\bf 67}, 1918 (1991);
T. Tohyama and S. Maekawa, Physica {\bf C 191}, 193 (1992);
E. Dagotto, F. Ortolani, and D. Scalapino, \phys {\bf 46}, 3183 (1992);
See also E. Dagotto, R. Joynt, A. Moreo, S. Bacci, and E.
Gagliano,
\prb 41, 9049, 1990.

\refis{super}
E. Dagotto, E., and J. Riera, \phys {\bf 46}, 12084 (1992).

\refis{emery87} For previous work on this phase using an attractive
V-term see
V. J. Emery, \lett {\bf 58}, 2794 (1987).

\refis{phsep}
V. J. Emery, S. A. Kivelson and H. Q. Lin, \lett {\bf 64},
475 (1990).

\refis{emery93}
V. J. Emery, and S. A. Kivelson, Physica {\bf C 209}, 597 (1993).

\refis{fano} For details on the technique used, we refer to
G. Fano, F. Ortolani, and F. Semeria, Int. J. Mod. Phys. B {\bf 3},
1845 (1990).

\refis{fujimori}
A. Fujimori, JJAP Series 7, Mechanisms of
Superconductivity p.p. 125 (1992); J. Phys. Chem. Solids, Vol. 53, 1595 (1992).

\refis{grilli}
This result has also been noticed in the context of the three
band Hubbard model. See for example
Grilli, M., R. Raimondi, C. Castellani, C. Di Castro, and G. Kotliar,
\lett {\bf 67}, 259 (1991);
R. Raimondi, C. Castellani, M. Grilli, Y. Bang, and G.
Kotliar, \phys {\bf 47}, 3331 (1993).

\refis{phsepexp}
P. C. Hammel, et al., \phys {\bf 42}, 6781 (1990);
Physica {\bf C 185}, 1095 (1991);
P. C. Hammel, E. Ahrens, A. Reyes, J. Thompson, Z. Fisk,
P. Canfield, J. Schirber and D. MacLaughlin, invited paper for workshop
on Phase Separation in Cuprate Superconductors, Erice, Italy (1992).

\refis{hellberg}
C. S. Hellberg, and E. J. Mele, \phys {\bf 48}, 646 (1993).




\refis{kivelson}
S. Kivelson, V. Emery, and H. Q. Lin, \phys {\bf 42}, 6523 (1990).


\refis{moreo}
A. Moreo, E. Dagotto, T. Jolicoeur and J. Riera,
\phys {\bf 42}, 6283 (1990).

\refis{moreo92}
A. Moreo, \phys {\bf 45}, 5059 (1992); and references therein.
See also M. Imada, and Y. Hatsugai, J. Phys. Soc. Jpn. {\bf 58},
3752 (1989), and D. J. Scalapino, S. R. White, and S. C. Zhang, \phys
{\bf 47}, 7995 (1993).

\refis{moreo93}
A. Moreo, FSU preprint, to appear in \phys, 1993;
See also R. R. P. Singh, and R. L. Glenister, \phys {\bf 46}, 11871 (1992).

\refis{ogata}
M. Ogata, M. Luchini, S. Sorella, and F. Assaad, \lett {\bf 66},
2388 (1991).

\refis{binding}
J. Bonca, P. Prelovsek, and I. Sega, \phys {\bf 39}, 7074 (1989);
J. Riera and A. P. Young, \prb 39, 9697, 1989;
Y. Hasegawa, and D. Poilblanc, \phys {\bf 40}, 9035 (1989);
E. Dagotto, J. Riera and A. P. Young, \prb 42, 2347, 1990;
M. Boninsegni and E. Manousakis, \phys {\bf 47}, 11897 (1993);
D. Poilblanc, J. Riera and E. Dagotto, FSU-SCRI preprint (1993).

\refis{qcd}
In this respect the results presented here are similar
in spirit to those obtained in the context of particle physics many
years ago, when it was shown that Quantum Chromodynamics (QCD)
confines at large bare couplings on a lattice
(see J. Kogut, Rev. Mod. Phys. {\bf 51}, 659 (1979); Rev. Mod. Phys. {\bf
55}, 775 (1983)).
Although this is not
at all the realistic regime, such a result was considered as
an important progress
towards showing that QCD confines. Actually, it was enough to carry out
a Monte Carlo simulation of lattice QCD
to show that the regime of strong coupling
and weak coupling (i.e. the physically
realistic regime due to asymptotic freedom) were
analytically connected. We hope that a similar program will start also
in the context of the 2D ${\rm t-J}$ model based on the results
discussed in this paper.

\refis{ds}
D. J. Scalapino, S. R. White, and S. C. Zhang, \lett
{\bf 68}, 2830 (1992).

\refis{scalettar}
R. T. Scalettar, et al., \phys {\bf 62}, 1407 (1989);
A. Moreo, and D. J. Scalapino, \lett {\bf 66}, 946 (1991).

\refis{sdw}
J. R. Schrieffer, X.-G. Wen, and S.-C. Zhang,
\phys {\bf 39}, 11663 (1989).

\refis{dwave}
Z.-X. Shen, et al., \lett {\bf 70}, 1553 (1993);
Hardy, W. N., D. A. Bonn, D. C. Morgan, R. Liang and
K. Zhang, submitted to PRL (1993).

\refis{sudbo2} A. Sudbo, S. Schmitt-Rink, and C. M. Varma, \phys
{\bf 46}, 5548 (1992).

\refis{sudbo}
A. Sudbo, C. M. Varma, T. Giamarchi, E. B. Stechel, and R. T.
Scalettar, \lett {\bf 70}, 978 (1993).


\refis{sano} K. Sano and Y. Ono, Physica {\bf C 205}, 170 (1993).
See also W. Bardford and M. W. Long, J. Phys. Cond. Matter {\bf 5},
199 (1993).

\refis{long} For a previous analysis changing the statistics of
particles in the ${\rm t-J}$ model see
M. W. Long and X. Zotos, \phys {\bf 45}, 9932 (1992).

\refis{troyer}
M. Troyer, H. Tsunetsugu, T. M. Rice, J. Riera, and E. Dagotto,
\phys {\bf 48}, 4002 (1993).

\refis{uchida}
S. Uchida, T. Ido, H. Takagi, T. Arima, Y. Tokura and
S. Tajima, \phys {\bf 43}, 7942 (1991);
D. B. Tanner, and T. Timusk, ``Optical
Properties of high temperature superconductors,''
{\it Physical} {\it Properties of} {\it High-Temperature
Superconductors III}, edited by Donald M. Ginsberg (World Scientific,
Singapore) pp 363--469 (1992).

\refis{varma}
C. M. Varma,
S. Schmitt-Rink, and E. Abrahams, Solid State Comm. {\bf
62}, 681 (1987).

\refis{singlet}
F. C. Zhang, and T. M. Rice, \phys {\bf 37}, 3759 (1988).

\endreferences

\endpage

\singlespace

\vskip 1cm

\bigskip
\centerline{\bf Figure Captions}
\medskip

\item{1.} a) Superconducting susceptibility $\chi^d_{sup}$ corresponding to
$\dx2y2$
correlations
(defined in the
text) as a function of ${\rm J/t}$ for the 2D ${\rm t-J}$ model, at
quarter-filling density. The results are obtained  on a $4 \times 4$
cluster; b) Pairing correlations $C({\bf m})$ as a function of distance
$m$, at ${\rm J/t=3}$ (i.e. where $\chi^d_{sup}$ peaks in Fig.1a), $\langle n
\rangle = 1/2$, using a $4 \times 4$ cluster. The full squares are
results for $\dx2y2$ symmetry, while the open squares correspond to
extended s-wave; c) Same as b) but
for different values of ${\rm J/t}$, and using only the $\dx2y2$
correlations.
The full squares correspond to
${\rm J/t=3}$, the triangles to ${\rm J/t=1}$, and the open squares to
${\rm J/t=4}$ (region of phase separation). These results are taken from
Dagotto and Riera, Phys. Rev. Lett. {\bf 70}, 682 (1993), and reproduced
here to make the paper self-consistent.

\item{2.} Phase diagram of the ${\rm t-J-V}$ model at quarter filling.
The full squares denote the positions of the peaks in $\chi^d_{sup}$
obtained using a $4 \times 4$ cluster. The other boundary of the
superconducting regime (dot-dashed line) should not be considered
quantitative, but only schematic.

\item{3.} a) Cluster of 20 sites used in this paper. The sites are
numbered such that the neighbors
can be identified once periodic
boundary conditions are applied. The four thick links represent dimers
formed in the large ${\rm V/t}$ limit as
discussed in the text; b) $\dx2y2$ pairing correlations (normalized
to one at distance zero), as a function of distance for ${\rm J/t=3}$,
and $\langle n \rangle = 1/2$, obtained on a $4 \times 4$ cluster (full
squares) and on a 20 site cluster (open squares).

\item{4.} Pairing correlation ${\rm C({\bf m})}$ as a function of
distance, for the 2D ${\rm t-J}$ model
on a 20 site cluster. The open triangles, full
triangles,
open squares, and full squares are results at couplings ${\rm J/t = 2.1}$,
${\rm
2.7}$, ${\rm 3.0}$, and ${\rm 3.5}$, respectively.
(a) corresponds to $\dx2y2$ symmetry and density $\langle n \rangle =
0.4$,
while (b) is for extended-s and the same density.
(c) and (d) are the same as (a) and (b), respectively,
but at a density $\langle n \rangle = 0.20$.

\item{5.} Results obtained using the Variational Monte Carlo technique
on a $8 \times 8$ cluster applied to the 2D ${\rm t-J}$ model,
as described in the text. GW corresponds to the energy of the
Gutzwiller state
Eq.(\call{s4}), while the energies of the
$s$ and $\dx2y2$ states are denoted by
``s-wave'' and ``d-wave'', respectively. The energy of the phase separated
state
is labeled PS. (a) corresponds to density $\langle n \rangle = 10/64
\approx 0.156$;
(b) $\langle n \rangle = 26/64 \approx 0.406 $; and
(c) $\langle n \rangle = 42/64 \approx 0.656 $;
(d) rough phase diagram of the 2D ${\rm t-J}$ model
predicted by the VMC approach.

\item{6.} (a) d-wave pairing susceptibility as a function of ${\rm J/t}$ for
the $4 \times 4$ cluster. The full squares correspond to $\langle n
\rangle = 0.50$, while the open squares are for $\langle n \rangle =
0.75$; (b) d-wave pairing correlations at ${\rm J/t=3}$ and
$\langle n \rangle = 0.50$ (full squares); and
${\rm J/t= 2}$ and $\langle n \rangle = 0.75$ (open squares);
(c) similar results as those shown in Fig.6a, but for the 1D chain at
${\rm J/t = 3}$.
Ful squares denote results at $\langle n \rangle = 0.50$;
open squares at $\langle n \rangle = 0.75$; and triangles
at $\langle n \rangle = 0.875$ ; (d) shows explicitly the pairing correlation
as
a function of distance for the 1D ${\rm t-J}$ chain, at ${\rm J/t=3}$.
The notation is as in Fig.6c.

\item{7.} Phase diagram of the 2D ${\rm t-J}$ model.
``d-wave'' denotes the
phase where the $\dx2y2$ correlations were found to be strong in the
numerical study presented in this paper. ``s-wave'' denotes the regime
where the VMC approach showed the presence of a stable s-wave
condensate. The dashed line separating d-wave from s-wave is schematic
since we only have results at a small number of electronic densities.
AF denotes the antiferromagnetic region close to half-filling. In this
regime we do not have enough accuracy in our analysis to complete the
phase diagram. The d-wave superconducting phase may or may not extend
into the small ${\rm J/t}$ region. Finally, ``PM'' denotes a
paramagnetic state.

\item{8.} Phase diagram of the ${\rm t-U-V}$ model in the mean-field
approximation, at half-filling. SDW denotes a spin-density-wave state,
CDW a charge-density-wave, PS corresponds to phase separation, while
SS and DS are superconducting states with $s$ and $\dx2y2$ symmetry,
respectively.

\item{9.} Ground state of the two band model for
different densities and couplings in the atomic limit. a),b) and c) are
in one dimension, while d) is in two dimensions. a)
half-filling ${\rm n_h = 0}$; b) ${\rm n_h = 2}$ and
${\rm V} < \Delta + {{3}\over{4}} {\rm U_p}$; c) ${\rm n_h = 2}$ and
${\rm V} > \Delta + {{3}\over{4}} {\rm U_p}$; d) ${\rm n_h = 2}$ and
large ${\rm V}$.

\item{10.} a) $\chi_{ps}$ (defined in the text) as a function of ${\rm
V}$, for ${\rm U_d = 7}$, ${\rm
U_p = 0}$, and $\Delta = 1.5$, on a 6 cell cluster
with 8 holes (i.e. ${\rm n_h = 2}$).
The open squares are results for ${\rm t_{pd} = 0.5}$,
while the full squares correspond to ${\rm t_{pd} = 1}$. The triangles
joined by dashed lines correspond to the order parameter X used
in Ref.(\cite{assaad}); (b) is the same as (a) but with 10 holes i.e.
${\rm n_h = 4}$.

\item{11.}
a) Pairing correlations ${\rm C(m)}$ (defined in the text)
as a function of distance
for the one dimensional ${\rm Cu - O}$ model with ${\rm n_h = 2}$,
at ${\rm U_d = 7}$, ${\rm
U_p = 0}$, $\Delta = 1.5$, and ${\rm t_{pd} = 1}$. The full triangles,
open circles, full squares, and open squares,
correspond to ${\rm V=0.0}$, ${\rm 2.0}$, ${\rm 4.0}$, and ${\rm 8.0}$,
respectively;
b) Same as Fig.11a but for ${\rm n_h = 4}$.

\item{12.} Schematic phase diagram of the two dimensional ${\rm t-J}$ model
when
the fermionic statistics of the particles is removed. FM denotes a
ferromagnetic region,  while PS is phase separation. In the narrow region
between these two phases, ``binding'' ( i.e. $\Delta_B < 0$) is
observed, but such a narrow strip may be a finite size effect (and thus
we included an interrogation
mark in the figure).

\endit